%% file: main.tex
\RequirePackage{fix-cm}
\documentclass[smallcondensed]{svjour3}
\smartqed  
\usepackage{graphicx}

\usepackage[breaklinks,colorlinks,citecolor=blue,linkcolor=blue]{hyperref}
\usepackage{etoolbox}

\usepackage[T1]{fontenc}
\usepackage[utf8]{inputenc}

\usepackage{booktabs}

\usepackage{graphicx}

\usepackage{natbib}

\def\makeheadbox{\relax}

\def\makeheadbox{{%
\hbox to0pt{\vbox{\baselineskip=10dd\hrule\hbox
to\hsize{\vrule\kern3pt\vbox{\kern3pt
\hbox{This is a pre-print of an article published in \textbf{Journal of Intelligent}}
\hbox{\textbf{Information Systems}. The final authenticated version is available online at:}
\hbox{\href{ https://doi.org/10.1007/s10844-019-00582-9 }{https://doi.org/10.1007/s10844-019-00582-9}.}
\kern3pt}\hfil\kern3pt\vrule}\hrule}%
\hss}}}

\begin{document}

\title{Machine Learning for Music Genre: Multifaceted Review and Experimentation with Audioset}

\author{Jaime Ram\'\i{}rez         \and
        M. Julia Flores
}

\institute{Jaime Ram\'\i{}rez \at
              Computing Systems Department, UCLM, Spain \\
              \email{jaime.ramirez@alu.uclm.es}           
           \and
           M. Julia Flores \at
              Computing Systems Department, UCLM, Spain \\
              \email{julia.flores@uclm.es}
}

\titlerunning{ML for music genre: multifaceted review and experimentation with Audioset}
\maketitle

\vspace{-20mm}

\begin{abstract}
Music genre classification is one of the sub-disciplines of music information retrieval (MIR) with growing popularity among researchers, mainly due to the already open challenges. Although research has been prolific in terms of number of published works, the topic still suffers from a problem in its foundations: there is no clear and formal definition of what genre is. Music categorizations are vague and unclear, suffering from human subjectivity and lack of agreement. In its first part, this paper offers a survey trying to cover the many different aspects of the matter. Its main goal is give the reader an overview of the history and the current state-of-the-art, exploring techniques and datasets used to the date, as well as identifying current challenges, such as this ambiguity of genre definitions or the introduction of human-centric approaches. The paper pays special attention to new trends in machine learning applied to the music annotation problem.  Finally, we also include a music genre classification experiment that compares different machine learning models using Audioset.

\keywords{Machine learning \and datasets \and music information retrieval \and classification algorithms \and music \and feed-forward neural networks.}
\CRclass{H.3 I.2.6 I.4 I.5 F.1 J.5}

\end{abstract}

\input{1-intro.tex}

\input{2-machine-learning.tex}

\input{3-models.tex}

\input{4-music-classification.tex}

\input{5-datasets.tex}

\input{6-experiments.tex}

\input{7-results.tex}

\input{8-conclusion.tex}

\begin{acknowledgements}
This work has been partially funded by FEDER funds and the Spanish Government (MICINN) through projects SBPLY/17/180501/000493 and TIN2016-77902-C3-1-P.
\end{acknowledgements}

\bibliographystyle{spbasic}

\bibliography{references.bib}

\end{document}

%% file: 1-intro.tex
\section{Introduction}

Music information retrieval (MIR) is an interdisciplinary field  which covers different aspects concerning the extraction of information from music \citep{downie2003music}, from sociological and musicological aspects to recommender systems, music generators or annotators \citep{liem2013musiclef, kitahara2017music}. Recent advances in machine learning (ML) models and  artificial intelligence (AI) have changed traditional approaches in MIR based sometimes on signal processing \citep{langkvist2014review} and generating more accurate results. 

One of the sub-problems of the music annotation domain  exploring these advances is music genre classification (MGC). Historically, acoustic and sound characteristics have been the main features to consider when performing genre classification. For example, Jazz is usually characterized by swing rhythms, improvisation and instruments such as piano and trumpet, whereas Electronic music can be identified by the use of synthesizers. It is reasonable to think that, if provided with raw signals and acoustic characteristics, genre classification could be more precise. However, this approach has been only partially effective. Although genre is a crucial descriptor, not only for MGC, but for many problems related, its definition is unclear. The categorization of a piece of music into a musical genre is often subject to different human perceptions, opinions and personal experiences, causing vague and fuzzy genre definitions \citep{aucouturier2003representing}. As a result, research does not show consensus in terms of what genre is, how taxonomies/categories are created or how these systems are evaluated. In fact, although existing categorizations or labellings are assigned and consumed by humans (experts or aficionados), the majority of these systems are hardly or not at all user-centered.

Like in other machine perception topics, the MGC research community has adopted advances in ML. In the last decade, we have witnessed a revolution of intelligent artificial perception systems and machine learning in general \citep{krizhevsky2012imagenet, xiong2018microsoft, he2016deep}. Since 2006, important discoveries have made neural networks gain popularity again, such as the ability to overcome a historic problem in this kind of models: the ability to train networks with many layers (deep neural networks) \citep{hinton2006fast}. This ability, combined with the increase of available labelled data, more computationally powerful hardware, and the easy access to these computing resources through cloud services, have brought us to a stage where most intelligent perception systems are now dominated by deep learning. These models are nowadays able to achieve outstanding results in artificial machine perception and cognition, e.g. paintings generation or speech recognition performed more accurately than humans. Although the deep learning approach has received substantial attention, proving to be the most powerful strategy in terms of perception problems, other approaches such as probabilistic modelling have also made big improvements in their application to perception problems, sometimes in combination with deep learning techniques \citep{wang2016towards, an2017naive}.

With no doubt, fields that have received interest lately are artificial vision, natural language processing, or speech recognition. The results achieved with deep neural networks (DNNs) for classification and regression problems in these areas have been remarkable  \citep{he2015delving, conneau2016very}. However, we believe that problems related to music, art or subjective perception currently offer more margin for research and improvement. To this end, we give a bird's eye view to the reader, by analyzing different historical and novel methods applied to MGC, identifying open challenges for future lines of research, presenting relevant publicly available datasets for the matter and finally experimenting with a collection of ML models trained with a recently published dataset: Audioset.

The reminder of the article is organized as follows. In Section II we explain how machine learning is applied to music annotation. Subsequently, in Section III, we introduce some ML methods suitable for music genre classification. In sections IV and V we analyze previous research, open challenges and available datasets for music genre classification. In sections VI and VII, we present our experiments and corresponding results. Finally, in section VIII, we conclude our study.

%% file: 2-machine-learning.tex
\section{Machine learning}
\label{sec:ml}

For the scope of this article we mainly cover supervised learning applied to music classification. Supervised learning constructs a hypothesis based on how the inputs of each sample relate to the corresponding expected output, with the aim of effectively predicting the output of new inputs. Problems where the output is numeric are called regression problems, whereas if the output is a class label (from a pre-defined set of possibilities), we have a classification problem.

There are also different types of classification problems depending on the number of the labels assigned to each sample. The most simple classification case is deciding whether a sample belongs to a given category, i.e., predicting if an audio clip is music. This is called binary classification, as there are only two possible labels, like a yes/no problem. More complex examples may need to distinguish between more than two classes, e.g., predicting the main music genre of a song. In this case, the classifier needs to decide between many classes (music genres). This is called multi-class classification. Lastly, we have multi-label classification, which happens when a sample is associated to multiple classes. For example, a song that presents a fusion of musical styles would normally be associated to multiple genre labels.

ML algorithms can be successfully applied to MIR problems. Particularly, when focusing on music classification problems, the approach follows a standard and general procedure. First, a group of key features relevant for the problem are extracted from the audio. Then, these features are fed into a model for training. Consequently, there are two distinguished phases: (1) Transformation of the audio clip into a set of representative features and (2) Construction of a model from these examples. The structure and nature of the model will be based on the paradigm used and the specific parametrizations for the training step. Once the training is done, we need an evaluation phase to assess how good the learned model is. This can be done in multiple ways. When the training dataset is balanced with respect to the class label, that is, there are approximately the same cases for every label, then accuracy is the most widely used measure. However, if there exist a class unbalance problem, accuracy cannot properly measure the quality of the model. This is because the rare class would have very little impact on accuracy as compared to the majority class. In that case, other metrics such as precision, recall or ROC area metrics are used. Most of these measures can be derived from the confusion matrix, which shows the correctly and incorrectly classified cases by class label.

As indicated before, MGC is usually a multi-label classification problem. Predictions for a song can result in a mixture of genres. From an evaluation perspective, samples can be fully correct, partially correct or fully incorrect, so we need ways to measure the performance of every predicted label. In order to cope with this, there are multiple techniques, such as instance techniques or ranking based metrics like average precision (AP) \citep{gibaja2015tutorial}.

%% file: 3-models.tex
\section{Models}

Among the most popular paradigms in ML we can find classification or decision trees, probabilistic graphical models, artificial neural networks (NNs) and support vector machines (SVMs). Besides, successful combinations of a set of models, known as ensembles, have been designed such as bagging, boosting and random forest.

\begin{figure}[!t]
\begin{center}
\centering
\includegraphics[width=0.99\linewidth,trim=0cm 0.2cm 0.25cm 0, clip]{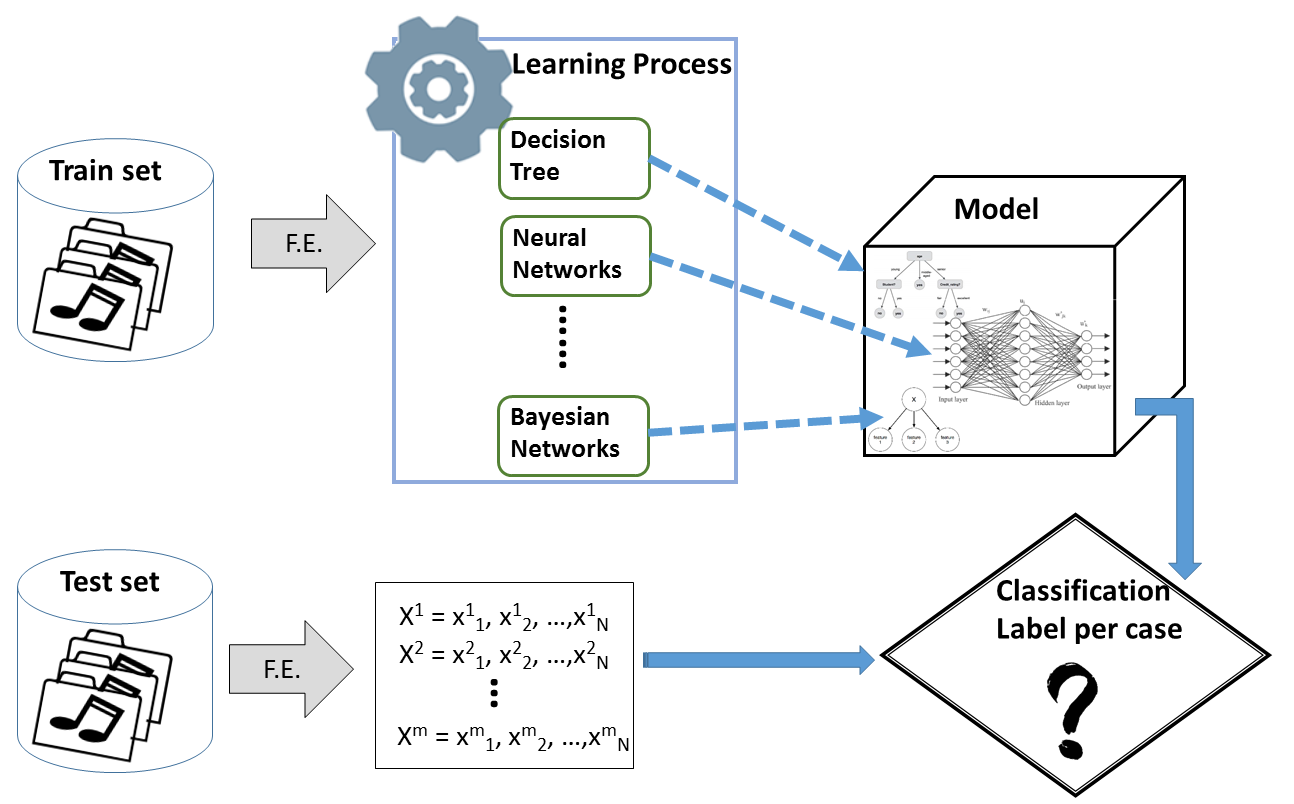}
\caption{Assuming we have previously done feature extraction (F.E.), the main task consists in learning a model from the training dataset, so that we can later classify new cases. For evaluation purposes, we will assess the model output using the cases in a test set.\label{fig:process}}
\end{center}
\end{figure}

Fig. \ref{fig:process} shows the general process, from clips we learn models which will help us predict new unknown clips.  Notice that Feature Extraction (F.E.) can be necessary in order to apply ML techniques. 

Among the training data, there will be one feature that is the objective, that is, the one to be automatically predicted. In our case music genre would take that role. Let us represent this class  as $Y$, and each possible class is a state of this variable; so if there are $k$ possible classes, then $Y=\{y_1,\dots,y_k\}$. For example $y_1$ could be jazz, $y_2$ opera, and so on. 

Classification consists in assigning one category (or value of the class variable) $y_l$ to a new object $x$, which is defined by the assignment of a set of values, $x = \{x_1,x_2,\cdots,x_n\}$, to the attributes $X_1,\dots,X_n$.  Per every clip in the dataset some features should have been extracted  earlier. This extraction + learning process could be done in a joint process, but there are some available datasets with this structure already available.

\subsection{Decision trees}

Decision trees are an attractive option for classification. They have been used in music classification \citep{basili2004classification, aucouturier2007signal}. A decision tree is a collection of decision nodes, connected by branches, extending downwards from the root node to leaf nodes. Beginning at the root, attributes are tested at the decision nodes, with each possible outcome resulting in a branch. Each branch then leads either to another decision node or to a terminating leaf node \citep{larose2014discovering}. These trees are able to depict understandable knowledge structures. Thus, it is easier to see which attributes play an important role in the decision process.

The algorithms for learning decision trees are diverse. One of the most simple and classic approaches is called ID3, which was improved by posterior methods C4.5 and C5.0 \citep{hssina2014comparative}. The construction of the tree uses an iterative process, where at each step the most discriminating node is chosen, and then each of the generated branches continue with a subset of the data coherent with that \emph{path}. The improvements on ID3 are mainly for speed and memory optimization, better tree expressiveness and lower overfitting. These algorithms are strongly based on an information measure, typically information gain ($G$), which is computed from entropy. %

\subsection{Probabilistic classifiers: Naive Bayes}

Probabilistic models have also been used in music analysis and retrieval \citep{paulus2009music, temperley2009unified, pickens2000comparison}, music recommendation \citep{park2006context, yoshii2008efficientRecommender}, music transcription \citep{benetos2015efficient} and music emotional perception \citep{egermann2013probabilistic, abdallah2002towards, wu2008probabilistic}. Naive Bayes (NB) classifier belongs to this family. It relies on the Bayes theorem and the assumption of independence between input features, given the class value. It is a \emph{naive} assumption where input features are assumed to contribute equally and independently to the target class. 

NB can be seen as a specific structure of a Bayesian Network specially intended for classification. When using NB as a classifier, emphasis is made on returning the right class variable, rather than numerically estimating the joint probability distribution in the most accurate way \citep{aguilera2011bayesian}.

NB has been proven comparable in performance with decision trees or SVM classifiers while showing better training speed than other models. Naive Bayes classifiers need a small amount of training data to provide good results.

\subsection{Support vector Machines (SVMs)}

Support Vector Machines try to determine the best and widest possible decision boundary between different classes \citep{burges1998tutorial}. They use vectors generating hyperplanes in high dimensional spaces that optimally separate points between different classes. SVMs learn these hyperplanes as decision boundaries. 

Optimal models choose the hyperplanes that maximize the margin between elements from different classes. This is the so-called large margin classification. Instances that mark the separation margin at both sides of the hyperplane are the support vectors. If the dataset is not linearly separable, more flexible models are used. Soft-margin classification tries to find a balance between the separation margin and the number of instances in the wrong side of the hyperplane. Other options are polynomial features or \textit{kernel} solutions.

\subsection{Fully Connected Neural Networks (FCNNs)}

Deep learning is the umbrella term that refers to different ML techniques based on deep artificial neural networks (ANNs). Inspired by their biological counterparts, they are made of basic units (neurons), with each neuron accepting several inputs that are combined (using an activation function) into a single output. This kind of classification model shows high accuracy, but the major disadvantage is the fact of being a \emph{black-box} model, which makes it difficult to be interpreted by humans.

Although ANNs dated back to the 1960s and the term \textit{Deep learning} had been around since the 80s \citep{dechter1986learning},   it was not until 2006 that the scientific community recovered the interest the topic, when \cite{hinton2006fast} managed to train ANNs with many neuron layers (deep networks). The possibility to train deep models allowed to learn different levels of knowledge for a problem, with the lower layers learning the low-level details, and the higher layers at the end learning more abstract features. These networks were coined as Fully Connected neural networks (FCNN) or simply Deep neural networks (DNNs), although they are also referred to as feed-forward networks or multilayer perceptrons (MLP).

Deep NNs have been used in many different areas, including audio/music problems \citep{deng2014deep}. Some of this research has yielded very successful results, e.g. speech recognition and synthesis \citep{hinton2012deep, amodei2016deep}, and there is a belief that other MIR problems have room for improvement with the potential of deep learning \citep{humphrey2013feature}.  Music analysis can combine different sources of information (audio signals, contextual information and human perception). This makes the problem richer and more suitable for deep models, which require bigger amounts of data. Different types of deep learning models have been applied to audio classification problems, such as  Deep Belief Networks (DBNs) and Restricted Boltzmann Machines (RBMs), Convolutional neural networks (CNNs) and Recurrent neural networks (RNNs).

\subsection{Convolutional neural networks (CNNs)}

CNNs are artificial NNs inspired by the distribution of neurons in the visual cortex of biological brains \citep{hubel1962receptive}. They are a type of multilayer perceptron (MLP), with some variations that make them specially suitable for artificial vision. CNNs use two types of layers: convolutional and pooling layers. Convolutional layers are built of neurons connected to their own receptive field. They work by applying a convolution operation over the input data. This convolution operation applies a filter matrix to the input data. Depending on the filter matrix, the convolution operation can detect certain shapes or features in the input data (like borders, horizontal lines, etc). Convolutional layers provide the network with the ability to learn lower-level information in the lower layers, while learning the details in the higher layers.  They are also capable of learning patterns in different places of the image. However, they are sensitive to the location of features in the input data. To fix this problem, pooling layers reshape the input to reduce the dimension (down-sample), making the network more robust to shifts in the position of features (location invariance) and additionally reducing big memory and computation loads. Usually, CNNs are built by stacking some pairs of convolutional and pooling layers, followed by a usual feed-forward network. Some CNNs architectures are AlexNet \citep{krizhevsky2012imagenet}, VGG \citep{simonyan2014very} and ResNet-50 \citep{he2016deep}.

CNNs have not only been applied to artificial vision. In audio-related problems, CNNs are used in combination with mid-level time-frequency image representation(spectrograms) as inputs. However, \cite{humphrey2013feature} question if the application of approaches successful in image recognition are suitable for audio signals, as time-series need another approach and better models \emph{ad-hoc} for audio need to be discovered. The same idea has also been explored for speech recognition \citep{zeghidour2018end}.

\subsection{Recurrent neural networks (RNNs)}

RNNs are a type of neural networks that connect their nodes forming a directed graph to represent a temporal sequence of events, which make them suitable for sequences of data, such as time series. In RNNs, each neuron behaves like a feed-forward neuron, except for an additional connection that sends the output back to itself. At each time step, a RNN neuron receives two vectors: the input vector and the output vector generated by itself from the previous time-step. This architecture allows the network to
preserve information along time. Sequences of data handled by RNNs do not need to have a fixed length, as RNNs can work with arbitrary lengths of data \citep{graves2012supervised, chung2014empirical}. Typical sequential scenarios are handwritten text, translation, speech recognition and music analysis. They are particularly suitable for the scope of this article as audio and music are temporal data of variable length.

When using Long Short-Term Memory, LSTM \citep{hochreiter1997long}, an improved RNN architecture, the model changes due to the use of memory cells that allow better gradient propagation over time, helping the network to learn relationships over longer sequences. Specifically, this architecture introduces the concepts of input gate, output gate and forget gate, and allows the network to use information beyond the immediately previous step (this is one drawback of traditional RNNs)  \citep{graves2013speech}. This design gives RNNs the ability to \emph{memorize} preceding events that allow them to remember previous states.

%% file: 4-music-classification.tex
\section{Music genre classification}

Since the seminal study of \cite{tzanetakis2002musical}, MGC has been a popular topic in the MIR community \citep{fu2011survey, knees2013survey, correa2016survey}. Despite the notable number of publications, about 500 according to \cite{sturm2012survey}, the field still presents open challenges nowadays, such as the ill-defined concept of genre, which is vague, fuzzy and subject to human perceptions \citep{aucouturier2003representing}, or the questionable significance of machine learning approaches centered on reproducing "ground truth" of given datasets \citep{sturm2014state}. This section summarizes the history of the topic to the date and covers these challenges.

Historically, research in MGC has been primarily addressed as a content-based problem in which learning musical structures from audio signals is key to infer a set of ground-truth labels that where categorized as genres \citep{aucouturier2003representing, panagakis2009music, fu2011survey}. 

\cite{tzanetakis2002musical} were the first to approach the problem. Working with 30-feature vectors extracted from audio signals, the authors achieved accuracy values of 61\% using a Gaussian mixture model (GMM) classifier and 60\% using a k-nearest neighbours (k-NN) classifier.  Their study included the creation of a dataset comprised of 1000 song clips evenly distributed across 10 genres. This collection, referred to as the GTZAN dataset, has become the most popular dataset for MGC research, although not the only one. Other datasets, reviewed in the next section, have received considerable attention too. In general, we will see that it is difficult to compare MGC studies due to the lack of standardization on the datasets and metrics used.

\subsection{Hand-crafted features}

After initial studies, research focused on exploring better feature extraction / selection methods and classifiers, in an effort to achieve better genre predictions and better audio labelling. Content-based features are extracted from raw signals to characterize audio in terms of pitch, timbre or rhythm, among others. Many of these features are extracted from short-time frames using the  Short Time Fourier Transform (STFT) \citep{rabiner1993fundamentals}. One of the most commonly used timbral features is \emph{Mel-frequency cepstral coefficients} (MFCCs), which captures short-term power spectrum of a sound, reducing high dimensional sound signals into low dimensional representations. MFCCs have been reported to perform well in different MIR domains \citep{logan2000mel, mandel2005song, tang2018music}, working as approximations of how the human auditory system maps frequencies to nerves in the inner ear. Another set of features used in research are spectral features, which give information about the sound spectrum. For example, \emph{spectral centroid} represents the centre of gravity of the magnitude spectrum of the SFTF and is correlated to the "brightness" of a sound \citep{tzanetakis2002musical}, whereas the \emph{spectral rolloff} measures the  spectral shape. \emph{Time domain Zero crosssings} inform about the noisiness of an audio signal \citep{tzanetakis2002musical}. \emph{Chroma} features, also popular, are a representation of pitch and enable modellings of melody and harmony, assuming  that humans perceive different pitches as similar if they are separated by an octave \citep{muller2015fundamentals}. 

\subsection{Automatic feature learning}

Even though substantial progress has been made with traditional approaches of feature selection based on hand-crafted signal processing techniques \citep{tzanetakis2002musical, li2003comparative, herrera2003automatic}, these approaches are still subject to the craftsmanship and engineering of signal processors coupled to the undertaken task, causing problems such as poor scalability or time-consumption \citep{humphrey2013feature}. A possible solution to this problem is to learn features automatically. This process, called audio feature learning, is often approached as an unsupervised learning problem and its is not restricted to genre classification. Music annotation, mood classification,  or instrument detection are good candidates for audio feature learning too \citep{langkvist2014review}.

Among techniques towards this end, we first find sparse coding (SC) algorithms. These approaches are particularly well suited to learn representations from audio as they mimic the way mammals auditory filters work \citep{olshausen1996emergence, smith2006efficient, lee2009unsupervised}. Although previous studies with SC techniques have shown notable results \citep{henaff2011unsupervised}, some of them have been reported by \cite{sturm2014state} to be no longer true, due to experimentation errors at \cite{panagakis2009music} and \cite{panagakis2010music} or irreproducible results \citep{chang2010music}. 

To cope with increasingly larger datasets and improve automatic feature selection, MIR researchers MIR researchers have adopted deep learning too \citep{sigtia2014improved}. \cite{humphrey2013feature} enumerate the benefits and obstacles of the application of deep learning to feature learning. In particular, learning features instead of hand-crafting them would improve scalability and foster the discovery of new features that could be otherwise missed in a manual selection processe. As obstacles, the authors mentioned the skepticism and practical difficulty to include deep learning in the toolbox.
However, in the last 5 years, the growth in popularity and acceptance of deep learning has been evident, not only in feature learning and MGC (as we will see in the remainder of the section), but in MIR and machine perception in general.

Application of deep architectures to MIR problems started with the deep belief network (DBN)  \citep{hinton2006fast}, which received considerable attention, specially between 2009 and 2013 \citep{lee2009unsupervised, hamel2010learning, schmidt2013learning}. In \citeyear{lee2009unsupervised},  \citeauthor{lee2009unsupervised} studied the use of convolutional DBNs for unsupervised feature learning in audio recognition tasks, such as genre or artist classification, outperforming MFCCs with 73.1\% accuracy on the ISMIR2004 Genre dataset\footnote{\url{https://www.upf.edu/web/mtg/ismir2004-genre}}. Shortly after, \cite{hamel2010learning} also chose DBNs to learn features automatically from music, outperforming MFCCs in tasks like MGC and tag discovery. The authors also reported problems with the training speed when experimenting with hyper-parameter tuning and the fact that features are inferred at frame level, so no temporal representation is learned, something that could have been interesting for music and audio. Later, in \citeyear{schmidt2013learning}, 
\citeauthor{schmidt2013learning} presented how DBNs can learn rhythm and melody representations for the prediction of music-related emotions \citep{schmidt2013learning}.

The success of CNNs in image classification has inspired MIR researchers to use visual representations of music audio (spectrograms). Using the short-time Fourier transform \citep{van2013deep}, spectrograms are extracted from audio and then fed into a CNN \citep{costa2017evaluation}. In an evaluation of CNNs for audio classification,\cite{wu2018reducing} observe that CNN models, concretely AlexNet, perform better than other deep architectures, such as  MLPs or RNNs, although they also report problems with the complexity of the AlexNet model. An example of feature learning with CNNs is Audioset \citep{jort2017audioset}. Using a transfer learning approach, the dataset provides pre-trained features that are the result of applying a VGG model to each sample \citep{simonyan2014very} and then extracting the output of the penultimate layer of the network. Other researchers have experimented with the direct use of raw audio signals as inputs for CNNs, getting results close to spectrogram-based approaches \citep{dieleman2014end}.

RNNs have also been considered to learn musical features, given their ability to operate on the linear progression of time. \cite{bock2016joint} use a combination of RNNs and Bayesian Networks to detect downbeats and upbeats. The RNN detects the beats, by activating any of their three output nodes (upbeat, downbeat, no beat), and the Bayesian network is subsequently used in a post-processing stage to infer tempo, phase, and metre.

\subsection{Content-based classification}

Progress in classification scores should not be fully accounted to advances in feature learning, as research have also evolved with regard to classification models.

The seminal work of \citeauthor{tzanetakis2002musical} used probabilistic and unsupervised approaches with GMM and k-NN classifiers. In particular, k-NN is a relatively popular model used for MGC. \cite{palmason2017competitiveness} report the best classification score for a k-NN model obtained in the GTZAN dataset, achieving accuracy scores in the range 80-81\%. \cite{iloga2018sequential} also employ k-NNs for MGC, extracting sequential patterns from music and generating music genre taxonomies. Other approaches such as granular computing have been used too \citep{ulaganathan2018granular}.

Soon after the work of \citeauthor*{tzanetakis2002musical}, SVMs gained popularity, and better accuracy scores were reported \citep{li2003comparative, henaff2011unsupervised}. Other probabilistic techniques would be later used such as probabilistic SVM outputs \citep{ness2009improving} or the Codeword Bernoulli Average model, which infers the probability of a tag appearing in a song \citep{hoffman2009easy}. \cite{silla2010improving} combine many different content-based features, using a genetic algorithm for the feature selection phase, and then use SVMs for genre classification.

More recently, deep learning models have been extensively employed for the matter. 
\cite{pons2016experimenting} classify samples of the Ballroom dataset experimenting with the dimensions of convolutional filters. In image processing, filters are spatial, whereas in audio spectrograms, filter dimensions are related to time and frequency. Therefore, filters can be selected to make the models more sensitive to temporal features (tempo, rhythm) or frequency patterns (instruments, timbre or equalization). The study shows results close to the baselines for the Ballroom dataset established by \cite{marchand2014modulation} and \cite{gouyon2004evaluating}. \cite{senac2017music} show that using a fine-tuned selection of CNNs input features, related to timbre, tonality and dynamics, could be more efficient and similar in accuracy than using spectrograms. \cite{tang2018music} propose the use of hierarchical LSTM models for MGC.

Researchers have also experimented with data combinations and model ensembles. \cite{nanni2016combining, nanni2018ensemble} explore the idea of merging visual and acoustic features for MGC. To this end, visual descriptors are extracted using a CNN and audio features are extracted with audio signal feature extraction methods. The combination of all features is finally fused to produce a classification with an SVM. \cite{kong2018audio} combine deep learning with the probability of each class, arranged in instance bags according to a multi-instance learning approach, obtaining a mean average precision (mAP) of 0.327 with Audioset. Probabilistic techniques combined with deep learning have been explored too. \cite{medhat2017masked} propose a type of neural networks designed for temporal signal recognition, the Conditional Neural Network and the Masked Conditional Neural Network achieving accuracy levels between 85\% and 86\% on the GTZAN and ISMIR2004 datasets.

\subsection{Open challenges}

Although genre is a crucial descriptor for many MIR related problems, there is not a generally agreed formal definition. Interpretations of music genre are often subject to different human perceptions, opinions and personal experiences, leading to different answers to the same question: \textbf{what is music genre?}. \cite{fabbri1999browsing} argues that not all genres are seen from the same perspective. Whereas some are subject to taxonomies defined in arbitrary dimensions (e.g. music technical features, industry categorizations), others can be more linked to cognitive psychological aspects. In fact, confusion is common between different taxonomy/categorization levels (should "Opera" and "Ethiopian Jazz" stay at different levels of a taxonomy?).

Examples of these discrepancies are varied. \cite{pachet2000taxonomy} discuss music genre in terms of taxonomies influenced by the music industry. \cite{burred2003hierarchical} also rely on taxonomies to approach music genre classification. \cite{celma2010music} highlights the connection of emotional, cultural and social aspects to music genre, and therefore the influence of personal experiences in genre categorizations.

Another source of confusion is the overlapping of different terms referring to music categorizations. It is common that genre and style overlap, albeit many \citep{fabbri1999browsing, moore2001categorical, i2009audio} agree in the definition of style as the way in which a musical piece is performed (how the music is played).

Others argue that approaching MGC just as a generic machine learning problem is not the most adequate strategy. \cite{sturm2014state} exposes the lack of a formal definition for MGR and questions whether systems that are able to reproduce the "ground truth" are really using relevant information to perform the prediction. This is corroborated by studies like the one by \cite{rodriguez2016analysing}, who demonstrate that classifiers systems benefit from very low sub-audible frequencies (how can we validate decisions based on inaudible sounds for us?). From the perspective of recommendation and personalization MIR systems, \cite{schedl2013neglected} expose, to our judgment, a particular important claim necessary for future progress: the need for user-centric approaches that help us cope with user properties and context (mood, past experiences, demographics, etc). 

Despite of these problematic aspects (unclear definitions, overlaps, ambiguity, subjectivity and in general, difficulty to gather ground truth), genre is a good tool to improve communication discussing musical categories. Users are accustomed to use genre as useful categorization for library/catalog browsing. Also categorizations of music affect music appreciation and cognition \citep{mckay2006musical}.

\subsection{Context-based and hybrid classification}

One of the possible approaches towards more human-centric approaches are context-based systems. Genre classification quality could be improved by not purely relying on the audio signal, but rather taking into consideration the context in which the music is played or additional sources of information associated to the piece \citep{knees2013survey}.

Context-based methods use information relative to the context in which a music piece occurs to establish a degree of similarity. To this end, high-level multimedia information is used, such as song lyrics \citep{mayer2008rhyme}, song tags \citep{levy2007semantic}, user ratings \citep{yang2012local}, artist biographies \citep{turnbull2009combining}, artist photographs \citep{libeks2011you} or co-occurrences in social media \citep{zangerle2012exploiting}. Most of these data are created by online communities and are accessible through web or mobile platforms. In the best cases, the data can be accessed by means of public APIs (Spotify, Last.fm). Despite of its potential, the downside of context data is its unbalanced nature, noise, and sparsity. Studies exploiting text information apply NLP techniques, such as TD-IDF, bag of words representation or Part-of-Speech tagging (PoS), although they suffer from limitations for languages other than English \citep{silla2010improving}. Methods which utilize user ratings to generate recommendations usually rely on matrix factorization techniques \citep{koenigstein2011yahoo}. 

Some authors include the temporal context in the sense of user tastes subject to temporal dynamics \citep{koenigstein2011yahoo} or events triggering a momentary inclination to listen to certain types of music \citep{yang2012local}. Previous work has also shown how to use hybrid methods that combine context and audio features for music and genre classification \citep{aucouturier2007signal, turnbull2009combining, mayer2008rhyme}. Other types of features have been added to the problem, such as in the study of \cite{prockup2015modeling}, who use features selected by music experts and extracted from the Music Genome Project\footnote{\url{http://www.pandora.com/about/mgp}}, or the work by \cite{han2010music}, in which the authors introduce emotion descriptors in a context-aware music recommender system.

\subsection{Working with human subjectivity}

In order to cope with the mentioned genre definition problems, It is interesting to review how musical features affect emotions and how ML models can cope with mood or subjectivity when classifying music \citep{yang2012machine}.

Some studies have explored the problem of mood prediction from music. \cite{schuller2010mister} present an SVM model to predict mood inspired by music. They mix audio features such as timbre and tone, with lyrics, chord sequences, music mode or dance style. \cite{schmidt2011learning} discuss how music emotions can be quantified and the difficulties caused by the subjectivity, ambiguity and perception of each listener. In order to approach this problem, they proposed a regression-based approach leveraging DBNs. They were able to learn features from audio that represent emotional associations between the acoustic content.

Lyrics can be an interesting source for emotions. \cite{an2017naive} approached music emotion perception from the point of view of text classification. By just using song lyrics as training data, they proposed a Bayesian classifier to infer the emotion caused by certain words or sentences.

Bayesian inference has also been used to model human perception of different types of art, including music. \cite{menendez2016towards} proposes a Bayesian brain theory that models art perception as the process of integrating current perceived art information with prior knowledge. In the case of music, the author considers music enjoyment as a prediction problem based on the music that the listener has consumed before, with the hypothesis that music enjoyment consists of predicting the sounds that come next and verifying whether these expectations are met \citep{meyer1957meaning}.

\cite{huang2017music} use emotions to find song fragments that represent the whole song. With CNNs, the authors use emotion recognition as a way to compare if a given part of a song corresponds to the chorus section. Their results show accurate emotion recognition, plus a way to detect the chorus part in music without prior information of the song structure.

%% file: 5-datasets.tex
\section{Data sets and sources}

Different taxonomies to model music genre have been created, but they do not show much consensus with regard to genre definitions and descriptors \citep{aucouturier2003representing}. As a consequence, datasets and data sources used in MGC research do not share a common structure, presenting different classes and a variable number or them. Moreover, there is not a commonly agreed standard dataset for the matter \citep{palmason2017competitiveness} and most of the published research use private datasets that do not allow to reproduce results \citep{sturm2012survey}.

Table \ref{table:datasets} lists datasets commonly used in MGC research, plus some with the potential to be used in future work. Below we give a more detailed description of each of them. For a more general MIR dataset structured list, we refer the reader to a community maintained repository\footnote{\url{https://github.com/ismir/mir-datasets}}.

\begin{table}[h!]
\setlength{\tabcolsep}{2pt}
\caption{Datasets for MGC, ordered by year of publication, specifying the size of the dataset, the number of classes or labels and the features each dataset provides.\label{table:datasets}}
\begin{center}
\begin{footnotesize}
\begin{tabular}{llllp{5.25cm}}\hline
\setlength{\arraycolsep}{0pt}
\textbf{Name}           & \textbf{Year} &  \textbf{Tags}     & \textbf{Size}               & \textbf{Features}    \\ 
\hline

GTZAN                   & 2002                     & 10                        & 1000                        & Audio \\
Ballroom                & 2004                     & 8                         & 698                         & Audio, tempo annotations \\
ISMIR2004               & 2004                     & 6                         & 729                         & Audio,  editorial metadata \\
Latin                   & 2008                     & 10                        & 3227                        & Extracted music vectors \\
Magnatagatune           & 2009                     & 188      & 25863                       & Audio, acoustic features  \\ 
MSD                     & 2011                     & 7643                      & 1000000                     & Acoustic features \\
 
MER31K                  & 2013                     & 190                       & 31427                       & Emotion tags                  \\  
AcousticBrainz (DB)     & 2015          &  -                         & 10469820\footnote{in 2018}            & Acoustic and high-level features \\
FMA                     & 2016                     & 163                       & 106574                      & Acoustic features, editorial metadata \\ 
Audioset                & 2017                     & 527                       & 2084320                     & VGG Embeddings, editorial metadata   \\ 
Kara1k                  & 2017                     & -                         & 1000                        & Audio, metadata, acoustic features \\ 

\end{tabular}
\end{footnotesize}
\end{center}
\end{table}

\subsection{GTZAN}

GTZAN was the first publicly available dataset for genre identification and was published as part of the seminal work in MGC by \cite{tzanetakis2002musical}. The dataset contains 1000 30-seconds long audio tracks, organized in ten popular music genres: Blues, Classical, Country, Disco, Hip Hop, Jazz, Metal, Pop, Reggae, and Rock. GTZAN has been used in many research papers \citep{henaff2011unsupervised, chang2010music, medhat2017masked, genussov2010musical, kotropoulos2010ensemble} and it is in fact the most used public dataset for music genre recognition research \citep{langkvist2014review}.
Given the popularity of the dataset, some works have assessed its quality and reliability for evaluating MGC models. \cite{sturm2012analysis} reports pitfalls, such a degree of duplication and mislabelling important enough to affect the results or research based on this dataset. \cite{palmason2017music} compare opinions from musical experts with GTZAN ground truths, obtaining an average agreement of 59.1\%. They conclude that better strategies are needed in the process of music genre tagging and propose the introduction of human ambiguity descriptors in dataset creation processes.

\subsection{ISMIR2004}
ISMIR2004 was created for the genre classification task in the ISMIR 2004 Audio Description Contest  \citep{cano2006ismir}. The collection is split in training, development and evaluation sets and the samples are grouped into 6 classes/genres: Classical, Electronic, Jazz/Blues, Metal/Punk, Rock/Pop and World. Each subset includes 729 tracks, arranged by class and unevenly distributed, with classical being the class with the highest number of samples and Jazz/Blues the lowest. Each sample contains the audio file for each track as well as some metadata such as artist, album and track name. According to \cite{sturm2012survey}, ISMIR2004 is the second most popular dataset in genre classification research.

\subsection{Latin music database}

The Latin music database was generated in \citeyear{silla2008latin} by  \citeauthor{silla2008latin} for MGC focused on Latin recordings. The dataset contains 3227 samples distributed across 10 Latin music genres:  Axé, Bachata, Bolero, Forró, Gaúcha, Merengue, Pagode, Salsa, Sertaneja and Tango. The collection was originally  compiled including original full recordings, but due to copyright reasons, only feature vectors extracted from the audio are available for download. Latin is the third most used dataset in genre classification research, according to \cite{sturm2012survey}.

\subsection{Ballroom}

Compiled in \citeyear{gouyon2004evaluating}, the Ballroom dataset by \citeauthor{gouyon2004evaluating} contains 698 audio samples with an approximate length of 30 seconds each associated to a genre class. There are eight possible classes, corresponding to ballroom musical genres, concretely, Jive, Quickstep, Tango, Waltz, Viennese Waltz, Samba, Cha Cha Cha and Rumba. 

In is the fourth most used dataset in MGC research \citep{sturm2012survey}.

\subsection{Million Song Dataset (MSD)}

The Million Song dataset is a freely-available dataset of audio features and meta-data extracted by means of The Echo Nest software API \footnote{\url{http://the.echonest.com/}}  \citep{mcfee2012million}. Among the features, we can find that each sample contains pitch, timbre and loudness. Its main purpose was to be a reference collection for research in MIR, while relieving researchers and practitioners from the hassle of creating large music datasets from scratch \citep{bertin2011millionSongDataset}.
The dataset is complemented with associated community-contributed collections such as the \textit{MusiXmatch} dataset for lyrics, Last.fm dataset for song tags and similarity, \textit{Taste profile} and \textit{thisismyjam-to-MSD} for user data and \textit{Tagtraum annotations/Top MAGD} for genre.

\subsection{Audioset}

Audioset is an ontology and human-labeled dataset of sounds extracted from YouTube \citep{jort2017audioset}. The associated dataset is currently formed by more than 2 million 10-second sound samples, annotated with 527 sound category labels. Although its initial purpose is not focused on music, the amount of genre labels and samples makes it a good candidate for MGC.

Audioset is available in two formats, csv and "embedding" representation format. The embeddings format delivers the dataset as TensorFlow record files, containing 128 features per second for each sample. These features are the result of the study carried out by 
\cite{hershey2017cnn}, who evaluate different CNN on YouTube-100M dataset for audio classification: fully connected, AlexNet \citep{krizhevsky2012imagenet}, VGG \citep{simonyan2014very}, Inception \citep{szegedy2016rethinking}, and ResNet-50 \citep{he2016deep}. Before this, most of these networks did not appear in publications applied to audio, being traditionally applied to the image recognition domain.

Audioset is one of the most recently created datasets. So far, efforts in research using it have been focused on evaluating how different deep learning models can perform in audio event recognition. \cite{xu2017surrey} employ AudioSet for weakly supervised audio event detection, whereas \cite{jansen2017towards} extract semantic representations from non-speech audio following an unsupervised approach. Since the dataset is intended for general purpose audio event classification, it is suitable for a variety of problems related to audio, such as music o video processing \citep{gao2018learning, zhou2018visual}.

\subsection{AcousticBrainz}

AcousticBrainz is an open project for gathering music information from users and making it available to the public. It includes low-level and high-level music information. Low-level information contains spectral, rhythm, time and tonal features, such as keys, scales, loudness, beats per minute. The high-level information is inferred by SVM models trained on the low-level features, and extracts information about genres, moods or instrumentation, among others. These models are trained on common datasets used in MIR, such as GTZAN \citep{tzanetakis2002musical}, Music Audio Benchmark Data Set, ISMIR2004 Rhythm Classification Dataset or MIREX Audio Mood Classification Dataset, and Music Technology Group datasets \citep{i2009audio, laurier2009music}. The authors remark the possibility of creating better annotated datasets to better infer high level features.

The project also includes a feature extractor and submission tool based on the Essentia library \citep{bogdanov2013essentia} that scans the user's music files and post the results to the AcoustricBrainz API \citep{porter2015acousticbrainz}. The project allows users to create their own datasets as subset of the genreal dataset, by annotating and sharing their datasets.

Contrary to other collections, AcousticBrainz is a live and on-growing database. Since its birth in 2014, the project has grown to more than ten million of data samples at the moment\footnote{July 2019. Statistics gathered from \url{https://acousticbrainz.org}}. To our knowledge, not much research has been conducted using data from AcousticBrainz \citep{bogdanov2017mediaeval}. It is indeed an open opportunity for research.

Recently, the same team of researchers have released the "AcousticBrainz Genre Dataset" \citep{bogdanov2019acousticbrainz}. Compiled specifically for MGC, using AcousticBrainz features, this dataset offers a combination of samples taken from four sources that offer genre categorizations of distinct nature: AllMusic, Discogs, Last.fm and Tagtraum. The objective of this dataset is to offer a large-scale, multifaceted and multi-label dataset for MGC.

\subsection{Freesound}
Freesound is an online database\footnote{\url{https://freesound.org}} of free audio excerpts \citep{font2013freesound}. In research, Freesound has been used in AudioPairBank, a study about how to analyze audio using verb and adjective tag pairs. This study is the application of a previous idea used with images for visual analysis \cite{borth2013large}. Freesound is extended by side projects such as Freesound Datasets: a collaborative platform for the creation of open datasets \citep{fonseca2017freesound}. The platform allows users to collaboratively annotate and create datasets.

\subsection{FMA}
FMA is a dataset created in 2016, especially intended for music analysis \citep{defferrard2016fma}. It includes 106574 full-length clips in high quality, accompanied by rich meta-data and pre-computed features. This fact provides practitioners with a wider range of feature combination possibilities. FMA is especially intended for MGC and possibly other MIR tasks, such as music recommendation. Yet, the dataset can present problems with its lack of mainstream music or genre diversity, as it only includes CC-licensed music. This can be problematic for domains aimed at the general public, e.g. music recommendation.

\subsection{MagnaTagATune}
The MagnaTagATune dataset\footnote{\url{http://mirg.city.ac.uk/codeapps/the-magnatagatune-dataset}} was constructed  in 2009 using data from TagATune\footnote{\url{http://tagatune.org/}}, an online game for collecting tags from music clips \citep{law2009input} and music from the Magnatune record label \citep{law2009evaluation}. It contains 25863 music clips distributed across 188 tags. It has been used in topics related to MGC \citep{mcfee2009heterogeneous, mandel2005song}.

\subsection{Other data}

There are some datasets and data sources that, although not originally intended for MGC, should be taken into account as they provide information significant for genre recognition, such as vocals or emotions. 

The Kara1k dataset \citep{bayle2017kara1k}, includes a variety of musical and audio features, as well as original and cover versions of the karaoke tracks recorded in professional studios. Among its features, it includes genre annotations, being an option for training MGC models focused on voice analysis.

Emotion is another dimension worth exploring in genre recognition. MoodSwings dataset offers information about the listener's mood \citep{schmidt2011modeling}. The dataset annotations were collected in the MoodSwings online game (no longer active), following the strategies used by MagnaTagATune \citep{law2009input} and Majorminer \citep{mandel2008web}.  The dataset includes 240 samples including acoustic features such as MFCCs, spectral and EchoNest audio features. A more recent dataset is MER31k \citep{yang2013quantitative}, created in 2013 specifically for Music emotion recognition (MER). The dataset annotates each of its 31427 samples across 190 mood-related tags, gathered from from Last.fm. 

Beat tracking is another area related to MGR, with extensive previous research and datasets \citep{li2003comparative,de2011corpus,bock2016joint}. As an example of beat tracking research closely related to genre, we find the work by \cite{hockman2012one}, in which the authors create a dataset for downbeat detection of fast paced electronic genres such as Hardcore, Jungle and Drum \& Bass, annotated by expert DJs.

Lastly, we find that the use of online services and APIs to gather information is common. Last.fm\footnote{\url{https://www.last.fm}} offers one of the most popular APIs to gather crowd-sourced tags from listeners. We can find some playlist collections, such as The Art of the Mix \citep{mcfee2011natural} and 8tracks \citep{bonnin2015automated}. Acoustic information can be obtained with the AcousticBrainz API and the EchoNest API (no longer available as it was acquired and absorbed by Spotify). AllMusic\footnote{\url{https://www.allmusic.com}} and Discogs\footnote{\url{https://www.discogs.com}} are the common sources for editorial metadata.

%% file: 6-experiments.tex
\section{Music genre classification with Audioset}

In order to gain more insights about how MGC works in practice, we have carried out a set of experiments to observe the performance of different models when provided with a labelled dataset. For our experiments, we use Audioset, due to its large collection of relevant music genre samples and also with the intention to open a line of research with this specific music collection. Whereas previous work with this novel dataset is addressed to general audio event detection, we focus on MGC.

The experiment consists in training and evaluating different ML models on a multi-label classification problem using a music genre subset of Audioset that we expressly compiled for the experiment. To explore the possibility of performance variation caused by unbalanced data, these models were trained on the Audioset balanced and unbalanced splits and later evaluated on the Audioset evaluation split.

\subsection{Experiments setup}

Audioset includes \textbf{52 music genre labels} that are present in almost 200000 records. The dataset provides pre-trained features extracted from a VGG-like CNN model, mapping each audio sample to vectors of numerical values, also called audio embeddings. This allows the researcher to feed these embeddings directly to the models. Additionally, the dataset also provides a balanced, unbalanced and evaluation splits that saves some dataset split work to the researcher.

In order to delimit the dataset to a MGC scope, we trimmed Audioset to include only samples tagged with any of the 52 music genre labels. The number of possible labels per sample was also reduced to 52 music genre labels. This reduction was applied to the balanced, unbalanced and evaluation sets. After this process, the balanced set was reduced from 22176 to 2490 samples, and the unbalanced set from 2042985 to 193481 samples, therefore leaving us with a music-genre subset that represents approximately a 10\% of the full Audioset.

 \subsection{Models}
For our experiments we selected Decision trees, NB classifiers, linear SVMs, DNNs, and RNNs. Previous work with Audioset uses deep learning techniques. However, to the best of our knowledge, there are no other studies evaluating other types of models, and we consider interesting to test how these models perform on the pre-trained audio embeddings provided by Audioset.

We have run a preliminary set of training sessions on the Audioset balanced split with 3-fold cross validation to select some of the best hyper-parameters for the Decision tree, the Naive Bayes classifier and the SVM. The same task has been carried out with the neural network models to select the number of layers and number of neurons per layer.
We found that tuning the hyper-parameters did not affect significantly to the performance of the experiments with Decision trees, NB classifiers nor Linear SVMs (especially the regularization parameter). In the case of deep learning models, the configuration of the network, in terms of number of layers and neurons, optimization algorithm and number or epochs, had more impact. The eventual configurations were:

\begin{itemize} 
\item The Decision Tree is limited to a max depth is 30.
\item The NB classifier assumes a normal prior distribution.
\item The SVM is linear.
\item The feed-forward NN uses 2 hidden layers with 768 units each, using batch normalization and dropout to prevent over-fitting. Adam optimizer \citep{kingma2014adam} with a learning rate of 0.001 is used to optimize the network parameters and binary cross entropy is used as loss function. The 2 hidden layers are composed of rectified linear units (ReLUs) \citep{nair2010rectified} instead of units using the sigmoid activation function. This leads to faster convergence and reduces the vanishing gradient problem, characteristic from traditional deep NNs.
\item The recurrent neural network uses LSTM 768 units in 1 hidden layer, using batch normalization and dropout to prevent over-fitting. The same optimizer and loss function used in the feed-forward NN is used with this one.
\end{itemize}

NB, SVM and Decision tree experiments have been developed with Scikit-learn \citep{pedregosa2011scikit}. For the NN and the RNN we have used Keras library \citep{chollet2015keras}. Data loading was performed with TensorFlow as Audioset is provided in TensorFlow record format (TFRecord format). In the case of Scikit-learn experiments, this format had to be converted to Numpy \citep{oliphant2006guide} format before running the experiments.
\subsection{Metrics}

In order to evaluate the experiments, we employed metrics used in the Google baseline \cite{jort2017audioset} and other Audioset related studies \citep{jort2017audioset, hershey2017cnn, gordon2018morphnet, kong2018audio}, namely mean Average Precision (mAP) and average area under the ROC Curve (AUC). In addition, we have also measured precision, recall and F1 score as it can help us to compare our work with previous work on music genre classification. Accuracy is not used as a metric due to the highly unbalanced nature of multi-label classification problem with a high number of labels. Lastly, training time in seconds is also measured to get an idea of which models are worthy in scenarios where the computational power and time are determining factors. The experiments have been executed in a Debian-based server with 8 Intel i7 cores, 1 Tesla K40c and 1 GeForce GTX 750.

%% file: 7-results.tex
\section{Results}

\begin{table}\setlength{\tabcolsep}{5pt}
    \centering
    \caption{Experiment results. Each of the models of the experiment is run in the \textbf{B}alanced and \textbf{U}nbalanced dataset, being the initial shown in column \textsf{T}. For each run, we calculated mean average precision ($\mu_{AP}$), mean ROC area under the curve ($\mu_{AUC}$), maximum average precision (max$_{AP}$),  maximum ROC area under the curve (max$_{AUC}$) and the training time in seconds. Maximum scores also indicate the genre obtaining the genre obtaining that score.  \label{table:1}}
    \resizebox{\textwidth}{!}{%
    \begin{tabular}{lllllll}
    \hline
    \textbf{Model} & \textbf{T} & \textbf{$\mu_{AP}$} & \textbf{$\mu_{AUC}$} & \textbf{max$_{AP}$}                 & \textbf{max$_{AUC}$}           & \textbf{Time (s)} \\ \hline
    Decision Tree  & B          & 0.060           & 0.560            & 0.170 (Music of Bollywood)     & 0.672 (Ambient music)     & 18          \\
    Decision Tree  & U        & 0.070           & 0.571            & 0.194 (Opera)                  & 0.661 (Ambient music)     & 1799.62                   \\
    Linear SVM     & B          & 0.126           & 0.596            & 0.432 (Music for children)     & 0.778 (Opera)             & 10.7          \\
    Linear SVM     & U        & 0.107           & 0.565            & 0.461 (Music for children)     & 0.900 (Music for children) & 29501                   \\
    Naive Bayes    & B          & 0.196           & 0.805            & 0.436 (Opera)                  & 0.913 (Beatboxing)        & 2.18           \\
    Naive Bayes    & U        & 0.176           & 0.830            & 0.358 (Electronic Dance Music) & 0.913 (New-age music)     & 100.7          \\
    NN             & B          & 0.417           & 0.909            & 0.779 (Music for children)     & 0.980 (Music for children) & 43.77                   \\
    NN             & U        & 0.465           & 0.930            & 0.820 (Music for children)     & 0.980 (A Capella)         & 767.486  \\   
    RNN            & B          & 0.356           & 0.890            & 0.699 (Opera)                  & 0.963 (Trance music)       & 282.63  \\  
    RNN            & U        & 0.437           & 0.929            & 0.794 (Music for children)     & 0.976 (Beatboxing)         & 5926.59  \\  
     
     \end{tabular}}
\end{table}

To give an idea of the performance of our experiments, we refer to the Audioset baseline audio event multi-label classifier \citep{jort2017audioset}, which uses a shallow fully-connected neural network. The baseline metric provided by Google for generic audio event classification with Audioset is a mean AP of 0.314 and an average AUC of 0.959, with a top performing class Music with AP of 0.896 and AUC of 0.951, Their worst AP was for “Rattle” with 0.020 and AUC of 0.796.

Our results are summarized in table \ref{table:1}. The Decision tree is the worst performing model with a mean AP of 0.070 and mean AUC of 0.570 on the unbalanced set, which is very close to random behaviour. Results on the tree trained on the balanced set are similar. The SVM classifier shows better results, especially in the balanced set, with a mean AP of 0.126 and a mean AUC of 0.596, higher than the Decision tree but still below the Google baseline. The SVM training phase on the unbalanced set is the slowest of all models. The Naive Bayes classifier stands out as the fastest model for training, with only 2.18 seconds required for the training phase on the balanced set, giving a mean AP / AUC of 0.196 / 0.805, and a mean AP / AUC of 0.176 / 0.830 on the unbalanced set. In terms of data balance, whereas the decision tree performs slightly better with unbalanced data, the SVM works better with balanced data. The NB classifier improves the AP with balanced data and shows better AUC with unbalanced data.

The neural network achieves a mean AP / AUC of 0.417 / 0.909 on the balanced dataset. The experiment with the unbalanced set improves these metrics with a mean AP of 0.465 and a mean AUC of 0.930. Performance shown by the neural network improves the baseline AP of 0.314. The RNN shows similar results, although not improving the performance of the NN. The unbalanced (and larger) dataset yields better scores in both networks.

\begin{figure*}[h!]
\setlength{\tabcolsep}{2pt}
\begin{center}
\begin{tabular}{ccc} \addtolength{\tabcolsep}{-7pt}
\includegraphics[width=0.33\linewidth,trim=0.26cm 0cm 0.3cm 0.1cm, clip]{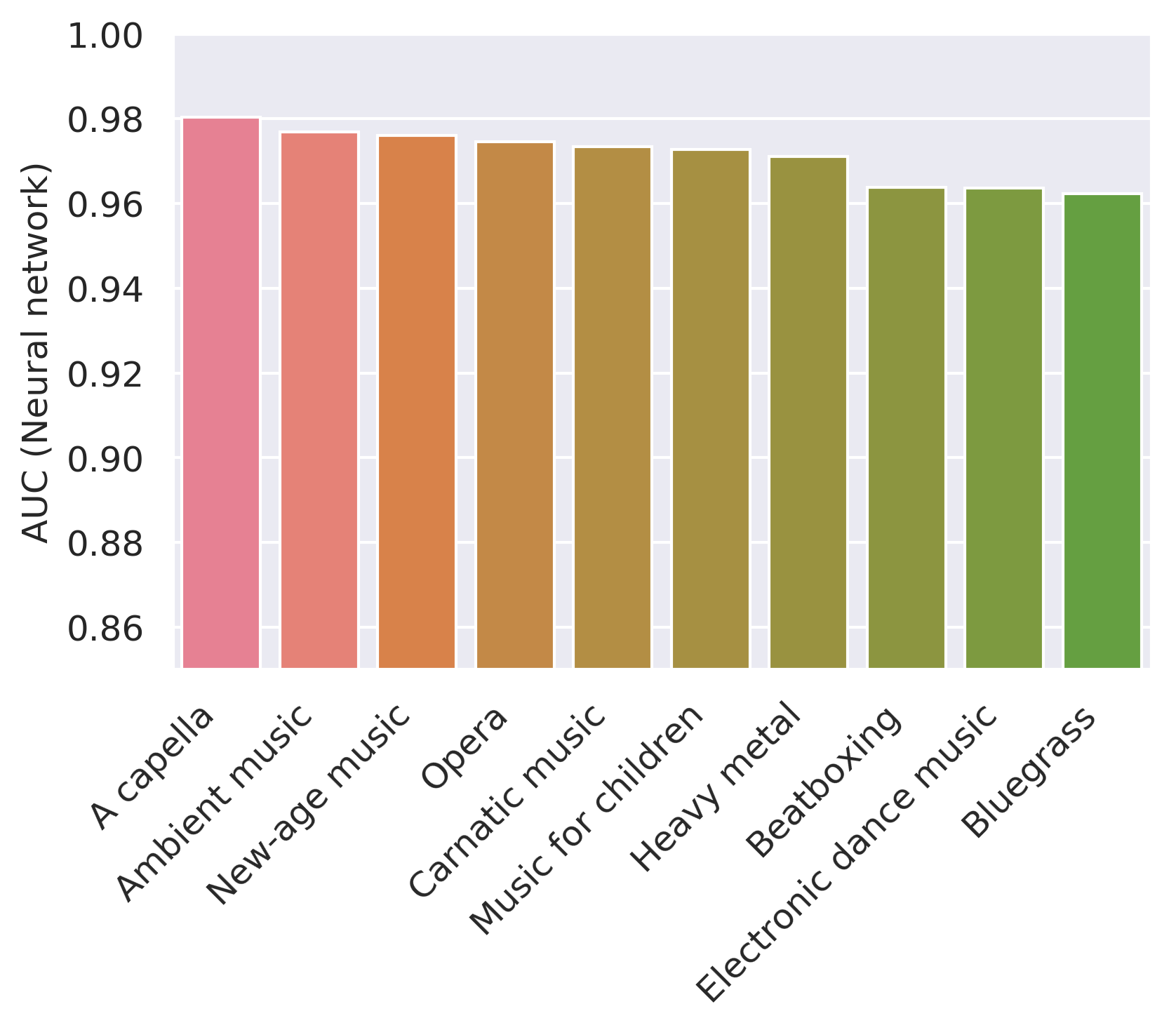}
& \includegraphics[width=0.33\linewidth,trim=0.26cm 0cm 0.3cm 0.1cm, clip]{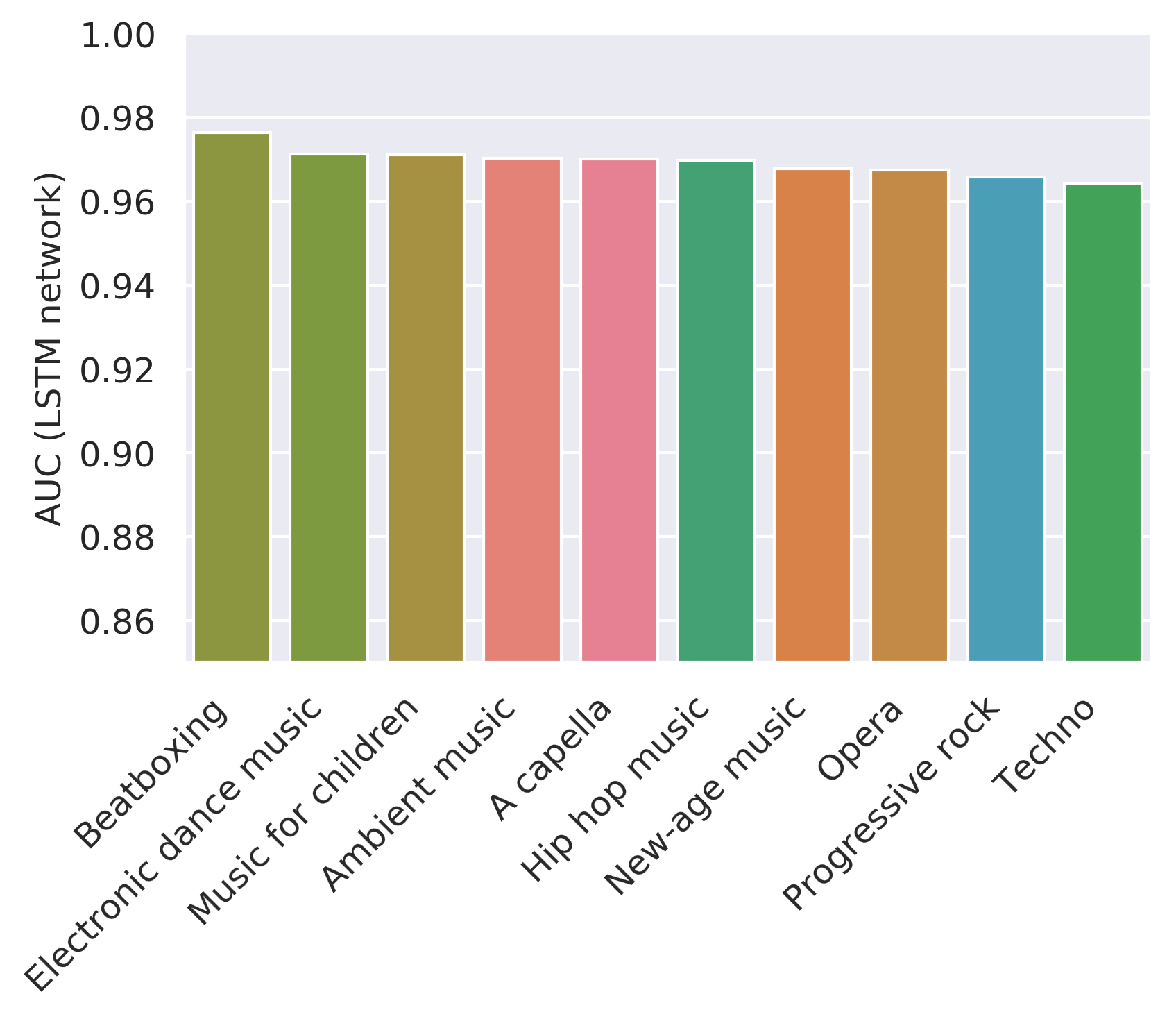}
& \includegraphics[width=0.33\linewidth,trim=0.26cm 0cm 0.3cm 0.1cm, clip]{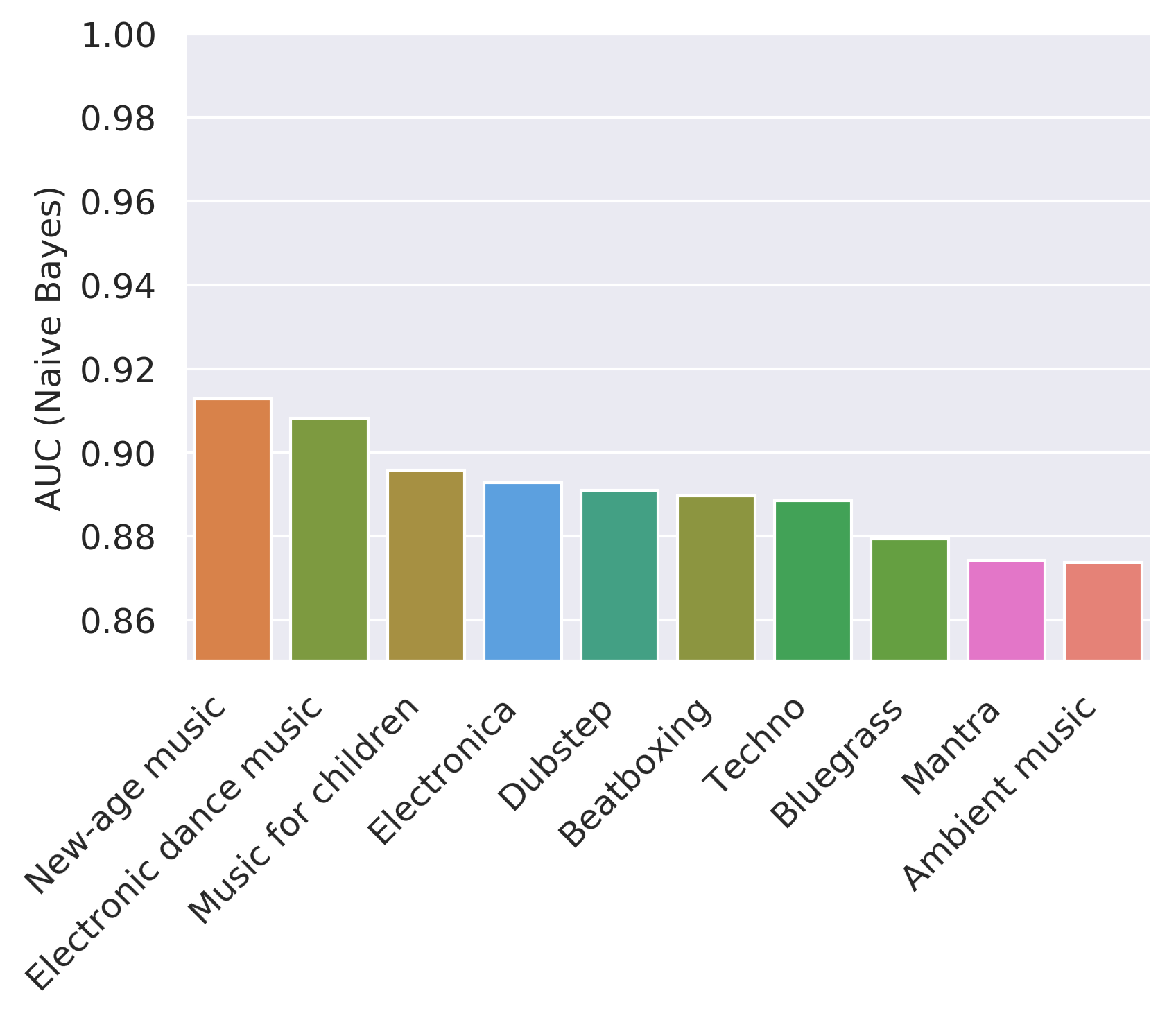}
\end{tabular}
\caption{Top 10 genre AUC scores for the neural network (left), recurrent LSTM network (middle) and Naive Bayes (right) classifiers.\label{fig:top_genres_auc_nn}}
\end{center}
\end{figure*}

\begin{figure*}[t]
\begin{center}
\includegraphics[width=1\linewidth, clip]{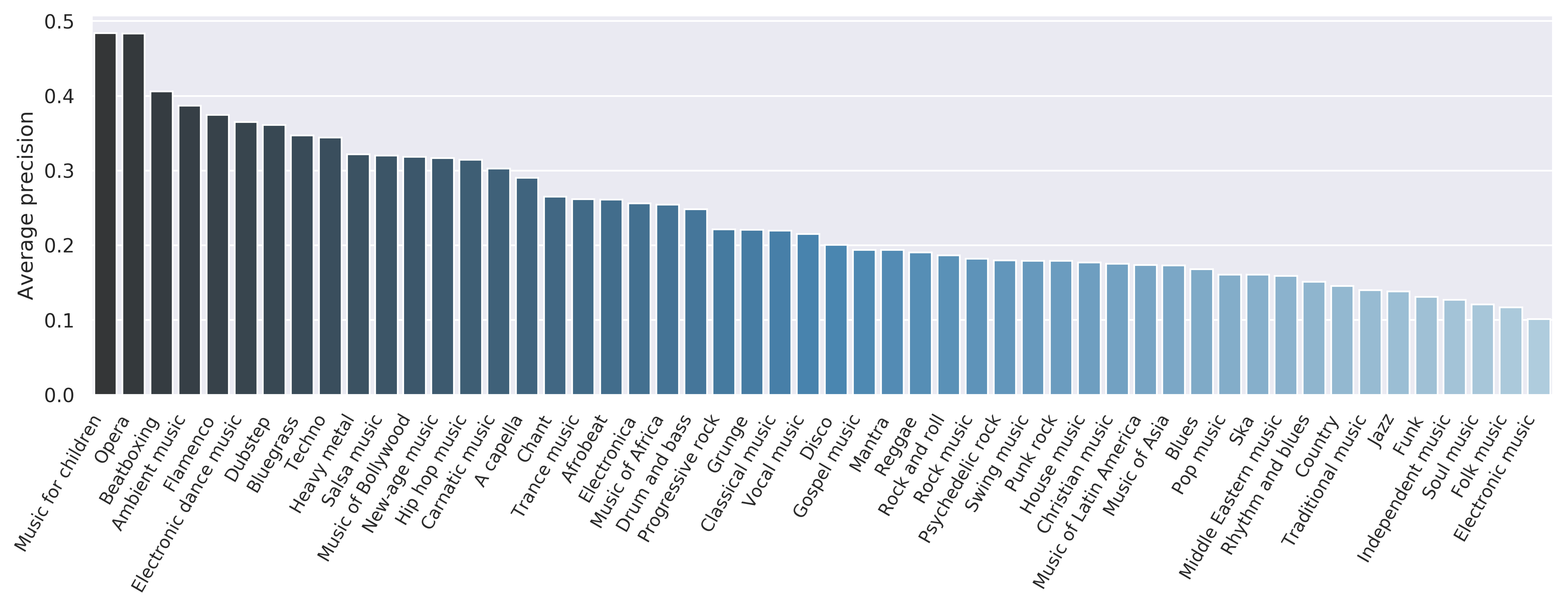}
\caption{Mean Average precision (AP) by genre for all models. This graph shows the average precision (AP) score for each music genre, averaged across the classifiers used in the experiment: decision tree, naive Bayes, SVM, neural network and recurrent (LSTM) neural network.\label{fig:best_classes_ap}}
\end{center}
\end{figure*}

\begin{figure}[t]

\begin{center}
\includegraphics[width=1\linewidth, clip]{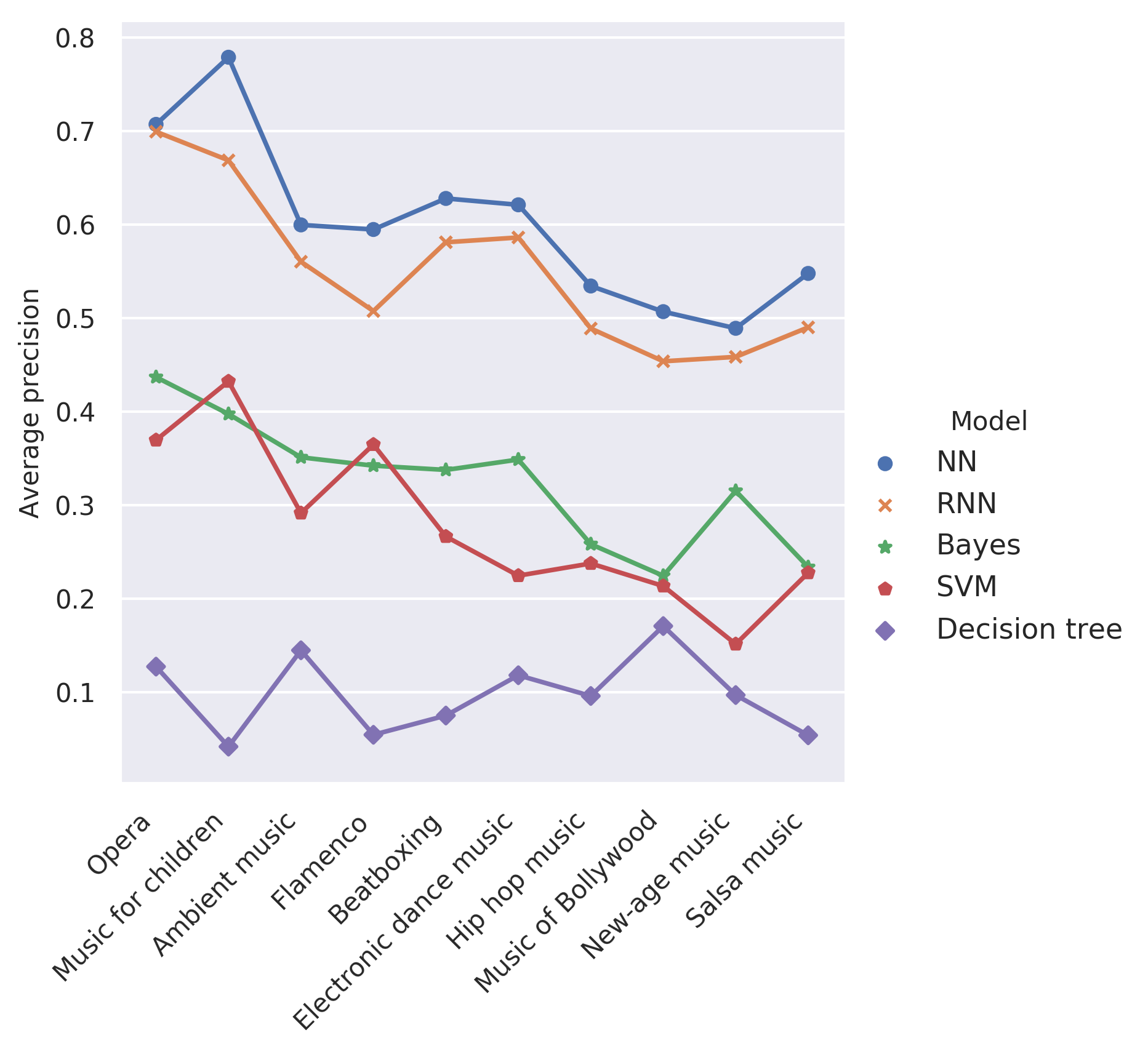}
\caption{Top 10 genre AUC scores by model for balanced Audioset. The genres are selected by averaging the AP metric across all models getting the first 10 results of the ordered resulting set.\label{fig:ap_genres_model}}
\end{center}
\end{figure}

Looking at the results of specific classes, both the SVM and the NB classifiers are capable of delivering better AP values than the Audioset baseline (Music for Children, Opera, Electronic Dance Music), and in the case of Naive Bayes, sometimes close to the AUC exhibited by deep learning models (New-age music, Beatboxing). Fig. \ref{fig:top_genres_auc_nn}(a) to (c) show that these classes clearly perform much better than others, and regardless of the model used, they usually rank high on the top-performing classes list. Fig. \ref{fig:best_classes_ap} shows best performing classes using AP as the metric. To identify best performing classes, we have averaged AP across all models and we have ranked genres based on this average. Fig. \ref{fig:ap_genres_model} shows the mean AUC performance for the top-10 classes when using the balanced Audioset split. The top ten has been calculated by adding the value of AP, AUC and F1.

\subsection{Results analysis}

Our experiments show AP and AUC scores for certain genres higher than the Google baseline. In the case of AP, the difference with the baseline is bigger, which can be caused by the fact that AP is directly correlated with the prior probability of each class. Therefore, it is natural that our AP scores tend to be higher, as only music genre classes have been considered in the experiment. 

The performance achieved using data from Audioset varies significantly between different models and music genres. Deep learning methods outperform the rest by a wide margin. In general, decision trees, Naive Bayes classifiers and SVMs performed worse than deep models, although Naive Bayes and SVMs have shown results closer to neural networks in certain classes, like "Music for children", or "Beatboxing" if we consider the AUC metric. In the case of the NB classifier, we have achieved AUC scores greater than 0.9 in those classes with less than 3 seconds for training with the data from the balanced split.

Let us remark the transfer learning nature of the experiments by using Audioset. Audioset provides features previously extracted from a VGG-like CNN model (embeddings). Looking at the data pipeline from an end-to-end perspective (from the raw signal to the final labels) our model is a CNN in which the first layers are provided by Audioset, followed by either NN, RNN, NB, SVM or decision tree classifiers. In the case of the NN, we have actually added more feed-forward layers to that CNN, thus following a standard CNN architecture. This is why we think the NN has yielded the best results.

%% file: 8-conclusion.tex
\section{Conclusions}

This paper contributes with up-to-date analysis of the existing research in music genre classification (MGC) from the main perspectives. We have first reviewed the typical workflow of MGC problem. Then, we have covered a wide spectrum of machine learning techniques and research related to the matter, as well as open challenges. We have also looked for the most relevant datasets in the field, indicating their main features and giving a thorough comparison among them. Finally, we have presented representative classifiers of the most spread paradigms, which have also been used in the experimental setup.

From our review, we can conclude that MGC is not an exception to deep learning trends in other machine perception areas, and has also widely adopted these models, both for feature learning and classification phases. The success of CNNs in artificial vision has influenced the way MIR problems are approached, using derived visual representations such as spectrograms of audio as input features.  We wonder, like other researchers have done, if better models are yet to be discovered, since CNNs are intended to handle images containing spatial data, not time and frequency dimensions.

With regard to challenges that remain open, the main problem is the lack of a formal definition for music genre. This leads to problems in the standardization and reproducibility of research, such as the lack of consensus in datasets, taxonomies or evaluation metrics. Moreover, the most used dataset for experimentation, GTZAN, has been reported to suffer from problems such as mislabelling and duplication. This fact can even question the results of previous research based on GTZAN. Better techniques are needed to create and standardize data sources for research, and these should include more user-centric descriptors about human subjectivity, judgment and uncertainty, as well as including music experts in the dataset creation process. From our perspective, we believe that probabilistic models could be a feasible approach to cope with this ambiguity.

With regard to our experiments with Audioset we have confirmed that neural networks work best and some interesting cases with other models have been discovered. The best performing model was the neural network, whereas the worst results were for the decision tree. Given the transfer learning nature of Audioset embeddings, extracted from a VGG-like CNN, it seems reasonable to think that adding feed-forward layers to the embeddings yields the best results. The NB classifier gives interesting results for some classes with AUC over 0.9 with just a few seconds of training, which leads us to think this model can be a promising option in scenarios with training time/computation constraints. In general, genres perform very differently. Various reasons can cause this: mislabeled samples, embeddings not capturing relevant information for those genres, or too broad labels. Mislabeled samples is something that the Audioset team is aware of. They provide supplementary files that inform about quality assessment a re-labelling status of samples. Unfortunately, we did not find any of the music genre samples in these files and therefore we did not have a metric of the quality of music genre labelling in Audioset. 

In the future we also plan to use other audio features extracted from the raw audio signals, as well as extending those features with other data sources such as context. Given the sample YouTube id, we can access not only the audio and video, but also user comments or tags, or even look for the song in other promising datasets for the matter, such as AcousticBrainz.

%% file: main.bbl
\begin{thebibliography}{148}
\providecommand{\natexlab}[1]{#1}
\providecommand{\url}[1]{{#1}}
\providecommand{\urlprefix}{URL }
\expandafter\ifx\csname urlstyle\endcsname\relax
  \providecommand{\doi}[1]{DOI~\discretionary{}{}{}#1}\else
  \providecommand{\doi}{DOI~\discretionary{}{}{}\begingroup
  \urlstyle{rm}\Url}\fi
\providecommand{\eprint}[2][]{\url{#2}}

\bibitem[{Abdallah(2002)}]{abdallah2002towards}
Abdallah SA (2002) Towards music perception by redundancy reduction and
  unsupervised learning in probabilistic models. PhD thesis, Queen Mary
  University of London

\bibitem[{Aguilera et~al.(2011)Aguilera, Fern{\'a}ndez, Fern{\'a}ndez,
  Rum{\'\i}, and Salmer{\'o}n}]{aguilera2011bayesian}
Aguilera P, Fern{\'a}ndez A, Fern{\'a}ndez R, Rum{\'\i} R, Salmer{\'o}n A
  (2011) Bayesian networks in environmental modelling. Environmental Modelling
  \& Software 26(12):1376--1388

\bibitem[{Amodei et~al.(2016)Amodei, Ananthanarayanan, Anubhai, Bai,
  Battenberg, Case, Casper, Catanzaro, Cheng, Chen et~al.}]{amodei2016deep}
Amodei D, Ananthanarayanan S, Anubhai R, Bai J, Battenberg E, Case C, Casper J,
  Catanzaro B, Cheng Q, Chen G, et~al. (2016) Deep speech 2: End-to-end speech
  recognition in english and mandarin. In: International Conference on Machine
  Learning, pp 173--182

\bibitem[{An et~al.(2017)An, Sun, and Wang}]{an2017naive}
An Y, Sun S, Wang S (2017) Naive bayes classifiers for music emotion
  classification based on lyrics. In: Computer and Information Science (ICIS),
  2017 IEEE/ACIS 16th International Conference on, IEEE, pp 635--638

\bibitem[{Aucouturier and Pachet(2003)}]{aucouturier2003representing}
Aucouturier JJ, Pachet F (2003) Representing musical genre: A state of the art.
  Journal of New Music Research 32(1):83--93

\bibitem[{Aucouturier et~al.(2007)Aucouturier, Pachet, Roy, and
  Beuriv{\'e}}]{aucouturier2007signal}
Aucouturier JJ, Pachet F, Roy P, Beuriv{\'e} A (2007) Signal + context= better
  classification. In: Proc. of the 8th International Conference on Music
  Information Retrieval, {ISMIR}, pp 425--430

\bibitem[{Basili et~al.(2004)Basili, Serafini, and
  Stellato}]{basili2004classification}
Basili R, Serafini A, Stellato A (2004) Classification of musical genre: a
  machine learning approach. In: ISMIR

\bibitem[{Bayle et~al.(2017)Bayle, Mar{\v{s}}{\'\i}k, Rusek, Robine, Hanna,
  Slaninov{\'a}, Martinovic, and Pokorn{\`y}}]{bayle2017kara1k}
Bayle Y, Mar{\v{s}}{\'\i}k L, Rusek M, Robine M, Hanna P, Slaninov{\'a} K,
  Martinovic J, Pokorn{\`y} J (2017) Kara1k: a karaoke dataset for cover song
  identification and singing voice analysis. In: Multimedia (ISM), 2017 IEEE
  International Symposium on, IEEE, pp 177--184

\bibitem[{Benetos and Weyde(2015)}]{benetos2015efficient}
Benetos E, Weyde T (2015) {An Efficient Temporally-Constrained Probabilistic
  Model for Multiple-Instrument Music Transcription}. In: Proc. of the 16th
  ISMIR Conference, pp 701--707

\bibitem[{Bertin-Mahieux et~al.(2011)Bertin-Mahieux, Ellis, Whitman, and
  Lamere}]{bertin2011millionSongDataset}
Bertin-Mahieux T, Ellis DP, Whitman B, Lamere P (2011) The million song
  dataset. In: Proc. of the 12th International Society for Music Information
  Retrieval Conference, {ISMIR}, pp 591--596

\bibitem[{B{\"o}ck et~al.(2016)B{\"o}ck, Krebs, and Widmer}]{bock2016joint}
B{\"o}ck S, Krebs F, Widmer G (2016) Joint beat and downbeat tracking with
  recurrent neural networks. In: ISMIR, pp 255--261

\bibitem[{Bogdanov et~al.(2013)Bogdanov, Wack, G{\'o}mez~Guti{\'e}rrez, Gulati,
  Herrera~Boyer, Mayor, Roma~Trepat, Salamon, Zapata~Gonz{\'a}lez, and
  Serra}]{bogdanov2013essentia}
Bogdanov D, Wack N, G{\'o}mez~Guti{\'e}rrez E, Gulati S, Herrera~Boyer P, Mayor
  O, Roma~Trepat G, Salamon J, Zapata~Gonz{\'a}lez JR, Serra X (2013) Essentia:
  An audio analysis library for music information retrieval. In: 14th
  Conference of the ISMIR; 2013. p. 493-8., ISMIR

\bibitem[{Bogdanov et~al.(2017)Bogdanov, Porter, Urbano, and
  Schreiber}]{bogdanov2017mediaeval}
Bogdanov D, Porter A, Urbano J, Schreiber H (2017) The mediaeval 2017
  acousticbrainz genre task: Content-based music genre recognition from
  multiple sources. In: Proc. of the MediaEval 2016 Workshop. Dublin, Ireland

\bibitem[{Bogdanov et~al.(2019)Bogdanov, Porter, Schreiber, Urbano, and
  Oramas}]{bogdanov2019acousticbrainz}
Bogdanov D, Porter A, Schreiber H, Urbano J, Oramas S (2019) The acousticbrainz
  genre dataset: Multi-source, multi-level, multi-label, and large-scale. In:
  20th International Society for Music Information Retrieval Conference (ISMIR
  2019), Delft, The Netherlands

\bibitem[{Bonnin and Jannach(2015)}]{bonnin2015automated}
Bonnin G, Jannach D (2015) Automated generation of music playlists: Survey and
  experiments. ACM Computing Surveys (CSUR) 47(2):26

\bibitem[{Borth et~al.(2013)Borth, Ji, Chen, Breuel, and
  Chang}]{borth2013large}
Borth D, Ji R, Chen T, Breuel T, Chang SF (2013) Large-scale visual sentiment
  ontology and detectors using adjective noun pairs. In: Proc. of the 21st ACM
  international conference on Multimedia, ACM, pp 223--232

\bibitem[{Burges(1998)}]{burges1998tutorial}
Burges CJ (1998) A tutorial on support vector machines for pattern recognition.
  Data mining and knowledge discovery 2(2):121--167

\bibitem[{Burred and Lerch(2003)}]{burred2003hierarchical}
Burred JJ, Lerch A (2003) A hierarchical approach to automatic musical genre
  classification. In: Proc. of the 6th international conference on digital
  audio effects, pp 8--11

\bibitem[{Cano et~al.(2006)Cano, G{\'o}mez~Guti{\'e}rrez, Gouyon,
  Herrera~Boyer, Koppenberger, Ong, Serra, Streich, and Wack}]{cano2006ismir}
Cano P, G{\'o}mez~Guti{\'e}rrez E, Gouyon F, Herrera~Boyer P, Koppenberger M,
  Ong BS, Serra X, Streich S, Wack N (2006) Ismir 2004 audio description
  contest. Tech. rep., Universitat Pompeu Fabra, Music technology Group

\bibitem[{Celma(2010)}]{celma2010music}
Celma O (2010) Music recommendation. In: Music recommendation and discovery,
  Springer, pp 43--85

\bibitem[{Chang et~al.(2010)Chang, Jang, and Iliopoulos}]{chang2010music}
Chang KK, Jang JSR, Iliopoulos CS (2010) Music genre classification via
  compressive sampling. In: ISMIR, pp 387--392

\bibitem[{Chollet et~al.(2015)}]{chollet2015keras}
Chollet F, et~al. (2015) Keras. \url{https://keras.io}

\bibitem[{Chung et~al.(2014)Chung, Gulcehre, Cho, and
  Bengio}]{chung2014empirical}
Chung J, Gulcehre C, Cho K, Bengio Y (2014) Empirical evaluation of gated
  recurrent neural networks on sequence modeling. arXiv preprint arXiv:14123555

\bibitem[{Conneau et~al.(2016)Conneau, Schwenk, Barrault, and
  LeCun}]{conneau2016very}
Conneau A, Schwenk H, Barrault L, LeCun Y (2016) Very deep convolutional
  networks for natural language processing. CoRR abs/1606.01781

\bibitem[{Corr{\^e}a and Rodrigues(2016)}]{correa2016survey}
Corr{\^e}a DC, Rodrigues FA (2016) A survey on symbolic data-based music genre
  classification. Expert Systems with Applications 60:190--210

\bibitem[{Costa et~al.(2017)Costa, Oliveira, and
  Silla~Jr}]{costa2017evaluation}
Costa YM, Oliveira LS, Silla~Jr CN (2017) An evaluation of convolutional neural
  networks for music classification using spectrograms. Applied soft computing
  52:28--38

\bibitem[{De~Clercq and Temperley(2011)}]{de2011corpus}
De~Clercq T, Temperley D (2011) A corpus analysis of rock harmony. Popular
  Music 30(1):47--70

\bibitem[{Dechter(1986)}]{dechter1986learning}
Dechter R (1986) Learning while searching in constraint-satisfaction problems.
  University of California, Computer Science Department, Cognitive Systems~…

\bibitem[{Defferrard et~al.(2017)Defferrard, Benzi, Vandergheynst, and
  Bresson}]{defferrard2016fma}
Defferrard M, Benzi K, Vandergheynst P, Bresson X (2017) {FMA:} {A} dataset for
  music analysis. In: Proc. of the 18th International Society for Music
  Information Retrieval Conference, {ISMIR} 2017, Suzhou, China, 2017, pp
  316--323

\bibitem[{Deng et~al.(2014)Deng, Yu et~al.}]{deng2014deep}
Deng L, Yu D, et~al. (2014) Deep learning: methods and applications.
  Foundations and Trends{\textregistered} in Signal Processing 7(3--4):197--387

\bibitem[{Dieleman and Schrauwen(2014)}]{dieleman2014end}
Dieleman S, Schrauwen B (2014) End-to-end learning for music audio. In: 2014
  IEEE ICASSP, IEEE, pp 6964--6968

\bibitem[{Downie(2003)}]{downie2003music}
Downie JS (2003) Music information retrieval. Annual review of information
  science and technology 37(1):295--340

\bibitem[{Egermann et~al.(2013)Egermann, Pearce, Wiggins, and
  McAdams}]{egermann2013probabilistic}
Egermann H, Pearce MT, Wiggins GA, McAdams S (2013) Probabilistic models of
  expectation violation predict psychophysiological emotional responses to live
  concert music. Cognitive, Affective, \& Behavioral Neuroscience
  13(3):533--553

\bibitem[{Fabbri(1999)}]{fabbri1999browsing}
Fabbri F (1999) Browsing music spaces: Categories and the musical mind. In:
  Proc. of Int. Association for the Study of Popular Music

\bibitem[{Fonseca et~al.(2017)Fonseca, Pons~Puig, Favory, Font~Corbera,
  Bogdanov, Ferraro, Oramas, Porter, and Serra}]{fonseca2017freesound}
Fonseca E, Pons~Puig J, Favory X, Font~Corbera F, Bogdanov D, Ferraro A, Oramas
  S, Porter A, Serra X (2017) Freesound datasets: a platform for the creation
  of open audio datasets. In: Proc. of the 18th ISMIR Conference; 2017. p.
  486-93., ISMIR

\bibitem[{Font et~al.(2013)Font, Roma, and Serra}]{font2013freesound}
Font F, Roma G, Serra X (2013) Freesound technical demo. In: Proc. of the 21st
  ACM international conference on Multimedia, ACM, pp 411--412

\bibitem[{Fu et~al.(2011)Fu, Lu, Ting, and Zhang}]{fu2011survey}
Fu Z, Lu G, Ting KM, Zhang D (2011) A survey of audio-based music
  classification and annotation. IEEE Trans on multimedia 13(2):303--319

\bibitem[{Gao et~al.(2018)Gao, Feris, and Grauman}]{gao2018learning}
Gao R, Feris R, Grauman K (2018) Learning to separate object sounds by watching
  unlabeled video. In: Proc. of the European Conference on Computer Vision
  (ECCV), pp 35--53

\bibitem[{Gemmeke et~al.(2017)Gemmeke, Ellis, Freedman, Jansen, Lawrence,
  Moore, Plakal, and Ritter}]{jort2017audioset}
Gemmeke JF, Ellis DPW, Freedman D, Jansen A, Lawrence W, Moore RC, Plakal M,
  Ritter M (2017) Audio set: An ontology and human-labeled dataset for audio
  events. In: Proc. of the {IEEE} International Conference on Acoustics, Speech
  and Signal Processing (ICASSP), pp 776--780

\bibitem[{Genussov and Cohen(2010)}]{genussov2010musical}
Genussov M, Cohen I (2010) Musical genre classification of audio signals using
  geometric methods. In: Signal Processing Conference, 2010 18th European,
  IEEE, pp 497--501

\bibitem[{Gibaja and Ventura(2015)}]{gibaja2015tutorial}
Gibaja E, Ventura S (2015) A tutorial on multilabel learning. ACM Computing
  Surveys (CSUR) 47(3):52

\bibitem[{Gordon et~al.(2018)Gordon, Eban, Nachum, Chen, Wu, Yang, and
  Choi}]{gordon2018morphnet}
Gordon A, Eban E, Nachum O, Chen B, Wu H, Yang TJ, Choi E (2018) Morphnet: Fast
  \& simple resource-constrained structure learning of deep networks. In: IEEE
  Conference on Computer Vision and Pattern Recognition (CVPR)

\bibitem[{Gouyon et~al.(2004)Gouyon, Dixon, Pampalk, and
  Widmer}]{gouyon2004evaluating}
Gouyon F, Dixon S, Pampalk E, Widmer G (2004) Evaluating rhythmic descriptors
  for musical genre classification. In: Proc. of the AES 25th International
  Conference, pp 196--204

\bibitem[{Graves(2012)}]{graves2012supervised}
Graves A (2012) Supervised sequence labelling. In: Supervised sequence
  labelling with recurrent neural networks, Springer, pp 5--13

\bibitem[{Graves et~al.(2013)Graves, Mohamed, and Hinton}]{graves2013speech}
Graves A, Mohamed Ar, Hinton G (2013) Speech recognition with deep recurrent
  neural networks. In: 2013 IEEE ICASSP, IEEE, pp 6645--6649

\bibitem[{Guaus(2009)}]{i2009audio}
Guaus E (2009) Audio content processing for automatic music genre
  classification: descriptors, databases, and classifiers. PhD thesis, PhD
  thesis, Universitat Pompeu Fabra, Barcelona, Spain

\bibitem[{Hamel and Eck(2010)}]{hamel2010learning}
Hamel P, Eck D (2010) Learning features from music audio with deep belief
  networks. In: ISMIR, Utrecht, The Netherlands, vol~10, pp 339--344

\bibitem[{Han et~al.(2010)Han, Rho, Jun, and Hwang}]{han2010music}
Han BJ, Rho S, Jun S, Hwang E (2010) Music emotion classification and
  context-based music recommendation. Multimedia Tools and Applications
  47(3):433--460

\bibitem[{He et~al.(2015)He, Zhang, Ren, and Sun}]{he2015delving}
He K, Zhang X, Ren S, Sun J (2015) Delving deep into rectifiers: Surpassing
  human-level performance on imagenet classification. In: Proc. of the IEEE
  international conference on computer vision, pp 1026--1034

\bibitem[{He et~al.(2016)He, Zhang, Ren, and Sun}]{he2016deep}
He K, Zhang X, Ren S, Sun J (2016) Deep residual learning for image
  recognition. In: Proc. of the IEEE conference on computer vision and pattern
  recognition (CVPR), pp 770--778

\bibitem[{Henaff et~al.(2011)Henaff, Jarrett, Kavukcuoglu, and
  LeCun}]{henaff2011unsupervised}
Henaff M, Jarrett K, Kavukcuoglu K, LeCun Y (2011) Unsupervised learning of
  sparse features for scalable audio classification. In: Proc. of the 12th
  International Society for Music Information Retrieval Conference, {ISMIR}, pp
  681--686

\bibitem[{Herrera-Boyer et~al.(2003)Herrera-Boyer, Peeters, and
  Dubnov}]{herrera2003automatic}
Herrera-Boyer P, Peeters G, Dubnov S (2003) Automatic classification of musical
  instrument sounds. J of New Music Research 32(1):3--21

\bibitem[{Hershey et~al.(2017)Hershey, Chaudhuri, Ellis, Gemmeke, Jansen,
  Moore, Plakal, Platt, Saurous, Seybold et~al.}]{hershey2017cnn}
Hershey S, Chaudhuri S, Ellis DP, Gemmeke JF, Jansen A, Moore RC, Plakal M,
  Platt D, Saurous RA, Seybold B, et~al. (2017) Cnn architectures for
  large-scale audio classification. In: Proc. of the {IEEE} International
  Conference on Acoustics, Speech and Signal Processing (ICASSP), pp 131--135

\bibitem[{Hinton et~al.(2012)Hinton, Deng, Yu, Dahl, Mohamed, Jaitly, Senior,
  Vanhoucke, Nguyen, Sainath et~al.}]{hinton2012deep}
Hinton G, Deng L, Yu D, Dahl GE, Mohamed Ar, Jaitly N, Senior A, Vanhoucke V,
  Nguyen P, Sainath TN, et~al. (2012) Deep neural networks for acoustic
  modeling in speech recognition: The shared views of four research groups.
  IEEE Signal Processing Magazine 29(6):82--97

\bibitem[{Hinton et~al.(2006)Hinton, Osindero, and Teh}]{hinton2006fast}
Hinton GE, Osindero S, Teh YW (2006) A fast learning algorithm for deep belief
  nets. Neural computation 18(7):1527--1554

\bibitem[{Hochreiter and Schmidhuber(1997)}]{hochreiter1997long}
Hochreiter S, Schmidhuber J (1997) Long short-term memory. Neural computation
  9(8):1735--1780

\bibitem[{Hockman et~al.(2012)Hockman, Davies, and Fujinaga}]{hockman2012one}
Hockman J, Davies ME, Fujinaga I (2012) One in the jungle: Downbeat detection
  in hardcore, jungle, and drum and bass. In: ISMIR, pp 169--174

\bibitem[{Hoffman et~al.(2009)Hoffman, Blei, and Cook}]{hoffman2009easy}
Hoffman MD, Blei DM, Cook PR (2009) Easy as cba: A simple probabilistic model
  for tagging music. In: ISMIR, vol~9, pp 369--374

\bibitem[{Hssina et~al.(2014)Hssina, Merbouha, Ezzikouri, and
  Erritali}]{hssina2014comparative}
Hssina B, Merbouha A, Ezzikouri H, Erritali M (2014) A comparative study of
  decision tree id3 and c4.5. International Journal of Advanced Computer
  Science and Applications(IJACSA), Special Issue on Advances in Vehicular Ad
  Hoc Networking and Applications 2014 4(2),
  \doi{10.14569/SpecialIssue.2014.040203}

\bibitem[{Huang et~al.(2017)Huang, Chou, and Yang}]{huang2017music}
Huang YS, Chou SY, Yang YH (2017) Music thumbnailing via neural attention
  modeling of music emotion. In: Proc. Asia Pacific Signal and Information
  Processing Association Annual Summit and Conference

\bibitem[{Hubel and Wiesel(1962)}]{hubel1962receptive}
Hubel DH, Wiesel TN (1962) Receptive fields, binocular interaction and
  functional architecture in the cat's visual cortex. The Journal of physiology
  160(1):106--154

\bibitem[{Humphrey et~al.(2013)Humphrey, Bello, and
  LeCun}]{humphrey2013feature}
Humphrey EJ, Bello JP, LeCun Y (2013) Feature learning and deep architectures:
  New directions for music informatics. J of Intelligent Information Systems
  41(3):461--481

\bibitem[{Iloga et~al.(2018)Iloga, Romain, and
  Tchuent{\'e}}]{iloga2018sequential}
Iloga S, Romain O, Tchuent{\'e} M (2018) A sequential pattern mining approach
  to design taxonomies for hierarchical music genre recognition. Pattern
  Analysis and Applications 21(2):363--380

\bibitem[{Jansen et~al.(2017)Jansen, Plakal, Pandya, Ellis, Hershey, Liu,
  Moore, and Saurous}]{jansen2017towards}
Jansen A, Plakal M, Pandya R, Ellis D, Hershey S, Liu J, Moore C, Saurous RA
  (2017) Towards learning semantic audio representations from unlabeled data.
  In: NIPS Workshop on Machine Learning for Audio Signal Processing (ML4Audio)

\bibitem[{Kingma and Ba(2015)}]{kingma2014adam}
Kingma DP, Ba J (2015) Adam: {A} method for stochastic optimization. In: Proc.
  of the 3rd International Conference on Learning Representations, {ICLR} 2015,
  San Diego, CA, USA, 2015

\bibitem[{Kitahara(2017)}]{kitahara2017music}
Kitahara T (2017) Music generation using bayesian networks. In: Altun Y, Das K,
  Mielik{\"a}inen T, Malerba D, Stefanowski J, Read J, {\v{Z}}itnik M, Ceci M,
  D{\v{z}}eroski S (eds) Machine Learning and Knowledge Discovery in Databases,
  Springer International Publishing, Cham, pp 368--372

\bibitem[{Knees and Schedl(2013)}]{knees2013survey}
Knees P, Schedl M (2013) A survey of music similarity and recommendation from
  music context data. ACM Trans on Multimedia Computing, Communications, and
  Applications (TOMM) 10(1):2

\bibitem[{Koenigstein et~al.(2011)Koenigstein, Dror, and
  Koren}]{koenigstein2011yahoo}
Koenigstein N, Dror G, Koren Y (2011) Yahoo! music recommendations: modeling
  music ratings with temporal dynamics and item taxonomy. In: Proc. of the 5th
  ACM conference on Recommender systems, ACM, pp 165--172

\bibitem[{Kong et~al.(2018)Kong, Xu, Wang, and Plumbley}]{kong2018audio}
Kong Q, Xu Y, Wang W, Plumbley MD (2018) Audio set classification with
  attention model: A probabilistic perspective. In: Proc. of the {IEEE}
  International Conference on Acoustics, Speech and Signal Processing (ICASSP),
  IEEE, pp 316--320

\bibitem[{Kotropoulos et~al.(2010)Kotropoulos, Arce, and
  Panagakis}]{kotropoulos2010ensemble}
Kotropoulos C, Arce GR, Panagakis Y (2010) Ensemble discriminant sparse
  projections applied to music genre classification. In: International
  conference on pattern recognition, IEEE, pp 822--825

\bibitem[{Krizhevsky et~al.(2012)Krizhevsky, Sutskever, and
  Hinton}]{krizhevsky2012imagenet}
Krizhevsky A, Sutskever I, Hinton GE (2012) Imagenet classification with deep
  convolutional neural networks. In: Advances in neural information processing
  systems, pp 1097--1105

\bibitem[{L{\"a}ngkvist et~al.(2014)L{\"a}ngkvist, Karlsson, and
  Loutfi}]{langkvist2014review}
L{\"a}ngkvist M, Karlsson L, Loutfi A (2014) A review of unsupervised feature
  learning and deep learning for time-series modeling. Pattern Recognition
  Letters 42:11--24

\bibitem[{Larose and Larose(2014)}]{larose2014discovering}
Larose DT, Larose CD (2014) Discovering knowledge in data: an introduction to
  data mining. John Wiley \& Sons

\bibitem[{Laurier et~al.(2009)Laurier, Meyers, Serra, Blech, and
  Herrera}]{laurier2009music}
Laurier C, Meyers O, Serra J, Blech M, Herrera P (2009) Music mood annotator
  design and integration. In: 7th International Workshop on Content-Based
  Multimedia Indexing. CBMI'09., IEEE, pp 156--161

\bibitem[{Law and Von~Ahn(2009)}]{law2009input}
Law E, Von~Ahn L (2009) Input-agreement: a new mechanism for collecting data
  using human computation games. In: Proc. of the SIGCHI Conference on Human
  Factors in Computing Systems, ACM, pp 1197--1206

\bibitem[{Law et~al.(2009)Law, West, Mandel, Bay, and
  Downie}]{law2009evaluation}
Law E, West K, Mandel MI, Bay M, Downie JS (2009) Evaluation of algorithms
  using games: The case of music tagging. In: ISMIR, pp 387--392

\bibitem[{Lee et~al.(2009)Lee, Pham, Largman, and Ng}]{lee2009unsupervised}
Lee H, Pham P, Largman Y, Ng AY (2009) Unsupervised feature learning for audio
  classification using convolutional deep belief networks. In: Advances in
  neural information processing systems, pp 1096--1104

\bibitem[{Levy and Sandler(2007)}]{levy2007semantic}
Levy M, Sandler M (2007) A semantic space for music derived from social tags.
  Austrian Compuer Society 1:12

\bibitem[{Li et~al.(2003)Li, Ogihara, and Li}]{li2003comparative}
Li T, Ogihara M, Li Q (2003) A comparative study on content-based music genre
  classification. In: Proc. of the 26th annual international ACM SIGIR
  conference on Research and development in informaion retrieval, ACM, pp
  282--289

\bibitem[{Libeks and Turnbull(2011)}]{libeks2011you}
Libeks J, Turnbull D (2011) You can judge an artist by an album cover: Using
  images for music annotation. IEEE MultiMedia 18(4):30--37

\bibitem[{Liem et~al.(2013)Liem, Orio, Peeters, and Schedl}]{liem2013musiclef}
Liem CCS, Orio N, Peeters G, Schedl M (2013) Musiclef 2013: Soundtrack
  selection for commercials. In: Proc. of the MediaEval 2013 Multimedia
  Benchmark Workshop, Barcelona, Spain, October 18-19, 2013.

\bibitem[{Logan et~al.(2000)}]{logan2000mel}
Logan B, et~al. (2000) Mel frequency cepstral coefficients for music modeling.
  In: ISMIR, vol 270, pp 1--11

\bibitem[{Mandel and Ellis(2005)}]{mandel2005song}
Mandel MI, Ellis D (2005) Song-level features and support vector machines for
  music classification. In: Proc. of the 6th International Conference on Music
  Information Retrieval, {ISMIR}, pp 594--599

\bibitem[{Mandel and Ellis(2008)}]{mandel2008web}
Mandel MI, Ellis DP (2008) A web-based game for collecting music metadata. J of
  New Music Research 37(2):151--165

\bibitem[{Marchand and Peeters(2014)}]{marchand2014modulation}
Marchand U, Peeters G (2014) The modulation scale spectrum and its application
  to rhythm-content description. In: Proc. of the 17th International Conference
  on Digital Audio Effects, pp 167--172

\bibitem[{Mayer et~al.(2008)Mayer, Neumayer, and Rauber}]{mayer2008rhyme}
Mayer R, Neumayer R, Rauber A (2008) Rhyme and style features for musical genre
  classification by song lyrics. In: Ismir, pp 337--342

\bibitem[{McFee and Lanckriet(2009)}]{mcfee2009heterogeneous}
McFee B, Lanckriet GR (2009) Heterogeneous embedding for subjective artist
  similarity. In: ISMIR, pp 513--518

\bibitem[{McFee and Lanckriet(2011)}]{mcfee2011natural}
McFee B, Lanckriet GR (2011) The natural language of playlists. In: ISMIR,
  vol~11, pp 537--542

\bibitem[{McFee et~al.(2012)McFee, Bertin-Mahieux, Ellis, and
  Lanckriet}]{mcfee2012million}
McFee B, Bertin-Mahieux T, Ellis DP, Lanckriet GR (2012) The million song
  dataset challenge. In: Proc. of the 21st International Conference on World
  Wide Web, ACM, pp 909--916

\bibitem[{McKay and Fujinaga(2006)}]{mckay2006musical}
McKay C, Fujinaga I (2006) Musical genre classification: Is it worth pursuing
  and how can it be improved? In: ISMIR, pp 101--106

\bibitem[{Medhat et~al.(2017)Medhat, Chesmore, and Robinson}]{medhat2017masked}
Medhat F, Chesmore D, Robinson J (2017) Masked conditional neural networks for
  audio classification. In: International Conference on Artificial Neural
  Networks, Springer, pp 349--358

\bibitem[{Menendez(2016)}]{menendez2016towards}
Menendez JA (2016) Towards a computational account of art cognition: unifying
  perception, visual art, and music through bayesian inference. Electronic
  Imaging 2016(16):1--10

\bibitem[{Meyer(1957)}]{meyer1957meaning}
Meyer LB (1957) Meaning in music and information theory. The Journal of
  Aesthetics and Art Criticism 15(4):412--424

\bibitem[{Moore(2001)}]{moore2001categorical}
Moore AF (2001) Categorical conventions in music discourse: Style and genre.
  Music and Letters 82(3):432--442

\bibitem[{M{\"u}ller(2015)}]{muller2015fundamentals}
M{\"u}ller M (2015) Fundamentals of music processing: Audio, analysis,
  algorithms, applications. Springer

\bibitem[{Nair and Hinton(2010)}]{nair2010rectified}
Nair V, Hinton GE (2010) Rectified linear units improve restricted boltzmann
  machines. In: Proc. of the 27th international conference on machine learning
  (ICML-10), pp 807--814

\bibitem[{Nanni et~al.(2016)Nanni, Costa, Lumini, Kim, and
  Baek}]{nanni2016combining}
Nanni L, Costa YM, Lumini A, Kim MY, Baek SR (2016) Combining visual and
  acoustic features for music genre classification. Expert Systems with
  Applications 45:108--117

\bibitem[{Nanni et~al.(2018)Nanni, Costa, Aguiar, Silla~Jr, and
  Brahnam}]{nanni2018ensemble}
Nanni L, Costa YM, Aguiar RL, Silla~Jr CN, Brahnam S (2018) Ensemble of deep
  learning, visual and acoustic features for music genre classification. J of
  New Music Research pp 1--15

\bibitem[{Ness et~al.(2009)Ness, Theocharis, Tzanetakis, and
  Martins}]{ness2009improving}
Ness SR, Theocharis A, Tzanetakis G, Martins LG (2009) Improving automatic
  music tag annotation using stacked generalization of probabilistic svm
  outputs. In: Proc. of the 17th ACM international conference on Multimedia, pp
  705--708

\bibitem[{Oliphant(2006)}]{oliphant2006guide}
Oliphant TE (2006) A guide to NumPy, vol~1. Trelgol Publishing USA

\bibitem[{Olshausen and Field(1996)}]{olshausen1996emergence}
Olshausen BA, Field DJ (1996) Emergence of simple-cell receptive field
  properties by learning a sparse code for natural images. Nature 381(6583):607

\bibitem[{Van~den Oord et~al.(2013)Van~den Oord, Dieleman, and
  Schrauwen}]{van2013deep}
Van~den Oord A, Dieleman S, Schrauwen B (2013) Deep content-based music
  recommendation. In: Advances in neural information processing systems, pp
  2643--2651

\bibitem[{Pachet and Cazaly(2000)}]{pachet2000taxonomy}
Pachet F, Cazaly D (2000) A taxonomy of musical genres. In: Content-Based
  Multimedia Information Access-Volume 2, pp 1238--1245

\bibitem[{P{\'a}lmason et~al.(2017{\natexlab{a}})P{\'a}lmason, J{\'o}nsson,
  Amsaleg, Schedl, and Knees}]{palmason2017competitiveness}
P{\'a}lmason H, J{\'o}nsson B{\TH}, Amsaleg L, Schedl M, Knees P
  (2017{\natexlab{a}}) On competitiveness of nearest-neighbor-based music
  classification: A methodological critique. In: International Conference on
  Similarity Search and Applications, Springer, pp 275--283

\bibitem[{P{\'a}lmason et~al.(2017{\natexlab{b}})P{\'a}lmason, J{\'o}nsson,
  Schedl, and Knees}]{palmason2017music}
P{\'a}lmason H, J{\'o}nsson B{\TH}, Schedl M, Knees P (2017{\natexlab{b}})
  Music genre classification revisited: An in-depth examination guided by music
  experts. Proc of CMMR

\bibitem[{Panagakis and Kotropoulos(2010)}]{panagakis2010music}
Panagakis Y, Kotropoulos C (2010) Music genre classification via topology
  preserving non-negative tensor factorization and sparse representations. In:
  Proc. of ICASSP, IEEE, pp 249--252

\bibitem[{Panagakis et~al.(2009)Panagakis, Kotropoulos, and
  Arce}]{panagakis2009music}
Panagakis Y, Kotropoulos C, Arce GR (2009) Music genre classification using
  locality preserving non-negative tensor factorization and sparse
  representations. In: Proc. of the 10th International Society for Music
  Information Retrieval Conference, {ISMIR}, pp 249--254

\bibitem[{Park et~al.(2006)Park, Yoo, and Cho}]{park2006context}
Park HS, Yoo JO, Cho SB (2006) A context-aware music recommendation system
  using fuzzy bayesian networks with utility theory. In: International
  conference on Fuzzy systems and knowledge discovery, Springer, pp 970--979

\bibitem[{Paulus and Klapuri(2009)}]{paulus2009music}
Paulus J, Klapuri A (2009) Music structure analysis using a probabilistic
  fitness measure and a greedy search algorithm. IEEE Trans on Audio, Speech,
  and Language Processing 17(6):1159--1170

\bibitem[{Pedregosa et~al.(2011)Pedregosa, Varoquaux, Gramfort, Michel,
  Thirion, Grisel, Blondel, Prettenhofer, Weiss, Dubourg
  et~al.}]{pedregosa2011scikit}
Pedregosa F, Varoquaux G, Gramfort A, Michel V, Thirion B, Grisel O, Blondel M,
  Prettenhofer P, Weiss R, Dubourg V, et~al. (2011) Scikit-learn: Machine
  learning in python. J of machine learning research 12(Oct):2825--2830

\bibitem[{Pickens(2000)}]{pickens2000comparison}
Pickens J (2000) A comparison of language modeling and probabilistic text
  information retrieval approaches to monophonic music retrieval. In: ISMIR

\bibitem[{Pons et~al.(2016)Pons, Lidy, and Serra}]{pons2016experimenting}
Pons J, Lidy T, Serra X (2016) Experimenting with musically motivated
  convolutional neural networks. In: 2016 14th International Workshop on
  Content-Based Multimedia Indexing (CBMI), pp 1--6,
  \doi{10.1109/CBMI.2016.7500246}

\bibitem[{Porter et~al.(2015)Porter, Bogdanov, Kaye, Tsukanov, and
  Serra}]{porter2015acousticbrainz}
Porter A, Bogdanov D, Kaye R, Tsukanov R, Serra X (2015) Acousticbrainz: a
  community platform for gathering music information obtained from audio. In:
  ISMIR Conference

\bibitem[{Prockup et~al.(2015)Prockup, Ehmann, Gouyon, Schmidt, Celma, and
  Kim}]{prockup2015modeling}
Prockup M, Ehmann AF, Gouyon F, Schmidt EM, Celma O, Kim YE (2015) Modeling
  genre with the music genome project: Comparing human-labeled attributes and
  audio features. In: ISMIR, pp 31--37

\bibitem[{Rabiner and Juang(1993)}]{rabiner1993fundamentals}
Rabiner LR, Juang BH (1993) Fundamentals of speech recognition, vol~14. PTR
  Prentice Hall Englewood Cliffs

\bibitem[{Rodr{\'{\i}}guez{-}Algarra et~al.(2016)Rodr{\'{\i}}guez{-}Algarra,
  Sturm, and Maruri{-}Aguilar}]{rodriguez2016analysing}
Rodr{\'{\i}}guez{-}Algarra F, Sturm BL, Maruri{-}Aguilar H (2016) Analysing
  scattering-based music content analysis systems: Where's the music? In: Proc.
  of the 17th International Society for Music Information Retrieval Conference,
  {ISMIR}, pp 344--350

\bibitem[{Schedl et~al.(2013)Schedl, Flexer, and Urbano}]{schedl2013neglected}
Schedl M, Flexer A, Urbano J (2013) The neglected user in music information
  retrieval research. Journal of Intelligent Information Systems 41(3):523--539

\bibitem[{Schmidt and Kim(2013)}]{schmidt2013learning}
Schmidt EM, Kim Y (2013) Learning rhythm and melody features with deep belief
  networks. In: ISMIR, pp 21--26

\bibitem[{Schmidt and Kim(2011{\natexlab{a}})}]{schmidt2011learning}
Schmidt EM, Kim YE (2011{\natexlab{a}}) Learning emotion-based acoustic
  features with deep belief networks. In: Applications of Signal Processing to
  Audio and Acoustics (WASPAA), 2011 IEEE Workshop on, IEEE, pp 65--68

\bibitem[{Schmidt and Kim(2011{\natexlab{b}})}]{schmidt2011modeling}
Schmidt EM, Kim YE (2011{\natexlab{b}}) Modeling musical emotion dynamics with
  conditional random fields. In: ISMIR, Miami (Florida), USA, pp 777--782

\bibitem[{Schuller et~al.(2010)Schuller, Hage, Schuller, and
  Rigoll}]{schuller2010mister}
Schuller B, Hage C, Schuller D, Rigoll G (2010) ‘mister dj, cheer me up!’:
  Musical and textual features for automatic mood classification. J of New
  Music Research 39(1):13--34

\bibitem[{Senac et~al.(2017)Senac, Pellegrini, Mouret, and
  Pinquier}]{senac2017music}
Senac C, Pellegrini T, Mouret F, Pinquier J (2017) Music feature maps with
  convolutional neural networks for music genre classification. In: Proc. of
  the 15th International Workshop on Content-Based Multimedia Indexing, ACM,
  p~19

\bibitem[{Sigtia and Dixon(2014)}]{sigtia2014improved}
Sigtia S, Dixon S (2014) Improved music feature learning with deep neural
  networks. In: 2014 IEEE ICASSP, IEEE, pp 6959--6963

\bibitem[{Silla et~al.(2010)Silla, Koerich, and Kaestner}]{silla2010improving}
Silla CN, Koerich AL, Kaestner CAA (2010) Improving automatic music genre
  classification with hybrid content-based feature vectors. In: Proc. of the
  2010 ACM Symposium on Applied Computing, ACM, pp 1702--1707

\bibitem[{Silla~Jr et~al.(2008)Silla~Jr, Koerich, and
  Kaestner}]{silla2008latin}
Silla~Jr CN, Koerich AL, Kaestner CA (2008) The latin music database. In:
  ISMIR, pp 451--456

\bibitem[{Simonyan and Zisserman(2015)}]{simonyan2014very}
Simonyan K, Zisserman A (2015) Very deep convolutional networks for large-scale
  image recognition. In: Proc. of 3rd International Conference on Learning
  Representations, {ICLR} 2015, San Diego, CA, USA, 2015

\bibitem[{Smith and Lewicki(2006)}]{smith2006efficient}
Smith EC, Lewicki MS (2006) Efficient auditory coding. Nature 439(7079):978

\bibitem[{Sturm(2012{\natexlab{a}})}]{sturm2012analysis}
Sturm BL (2012{\natexlab{a}}) An analysis of the gtzan music genre dataset. In:
  Proc. of the 2nd international ACM workshop on Music information retrieval
  with user-centered and multimodal strategies, ACM, pp 7--12

\bibitem[{Sturm(2012{\natexlab{b}})}]{sturm2012survey}
Sturm BL (2012{\natexlab{b}}) A survey of evaluation in music genre
  recognition. In: International Workshop on Adaptive Multimedia Retrieval,
  Springer, pp 29--66

\bibitem[{Sturm(2014)}]{sturm2014state}
Sturm BL (2014) The state of the art ten years after a state of the art: Future
  research in music information retrieval. Journal of New Music Research
  43(2):147--172

\bibitem[{Szegedy et~al.(2016)Szegedy, Vanhoucke, Ioffe, Shlens, and
  Wojna}]{szegedy2016rethinking}
Szegedy C, Vanhoucke V, Ioffe S, Shlens J, Wojna Z (2016) Rethinking the
  inception architecture for computer vision. In: Proc. of the IEEE Conference
  on Computer Vision and Pattern Recognition, pp 2818--2826

\bibitem[{Tang et~al.(2018)Tang, Chui, Yu, Zeng, Wong et~al.}]{tang2018music}
Tang CP, Chui KL, Yu YK, Zeng Z, Wong KH, et~al. (2018) Music genre
  classification using a hierarchical long short term memory (lstm) model. In:
  Proc. of the 3rd International Workshop on Pattern Recognition

\bibitem[{Temperley(2009)}]{temperley2009unified}
Temperley D (2009) A unified probabilistic model for polyphonic music analysis.
  J of New Music Research 38(1):3--18

\bibitem[{Turnbull et~al.(2009)Turnbull, Barrington, Lanckriet, and
  Yazdani}]{turnbull2009combining}
Turnbull DR, Barrington L, Lanckriet G, Yazdani M (2009) Combining audio
  content and social context for semantic music discovery. In: Proc. of the
  32nd international ACM SIGIR conference on Research and development in
  information retrieval, ACM, pp 387--394

\bibitem[{Tzanetakis and Cook(2002)}]{tzanetakis2002musical}
Tzanetakis G, Cook P (2002) Musical genre classification of audio signals. IEEE
  Trans on speech and audio processing 10(5):293--302

\bibitem[{Ulaganathan and Ramanna(2018)}]{ulaganathan2018granular}
Ulaganathan AS, Ramanna S (2018) Granular methods in automatic music genre
  classification: a case study. J of Intelligent Information Systems pp 1--21

\bibitem[{Wang and Yeung(2016)}]{wang2016towards}
Wang H, Yeung DY (2016) Towards bayesian deep learning: A framework and some
  existing methods. IEEE Trans on Knowledge and Data Engineering
  28(12):3395--3408

\bibitem[{Wu and Jeng(2008)}]{wu2008probabilistic}
Wu TL, Jeng SK (2008) Probabilistic estimation of a novel music emotion model.
  In: International Conference on Multimedia Modeling, Springer, pp 487--497

\bibitem[{Wu and Lee(2018)}]{wu2018reducing}
Wu Y, Lee T (2018) Reducing model complexity for dnn based large-scale audio
  classification. In: Proc. of the {IEEE} International Conference on
  Acoustics, Speech and Signal Processing (ICASSP), pp 331--335

\bibitem[{Xiong et~al.(2018)Xiong, Wu, Alleva, Droppo, Huang, and
  Stolcke}]{xiong2018microsoft}
Xiong W, Wu L, Alleva F, Droppo J, Huang X, Stolcke A (2018) The microsoft 2017
  conversational speech recognition system. In: Proc. of the IEEE international
  conference on acoustics, speech and signal processing (ICASSP), IEEE, pp
  5934--5938

\bibitem[{Xu et~al.(2017)Xu, Kong, Wang, and Plumbley}]{xu2017surrey}
Xu Y, Kong Q, Wang W, Plumbley MD (2017) Surrey-cvssp system for {DCASE2017}
  challenge task4. arXiv preprint arXiv:170900551

\bibitem[{Yang et~al.(2012)Yang, Chen, Zhang, Lu, and Yu}]{yang2012local}
Yang D, Chen T, Zhang W, Lu Q, Yu Y (2012) Local implicit feedback mining for
  music recommendation. In: Proc. of the 6th ACM conference on Recommender
  systems, ACM, pp 91--98

\bibitem[{Yang and Chen(2012)}]{yang2012machine}
Yang YH, Chen HH (2012) Machine recognition of music emotion: A review. ACM
  Trans on Intelligent Systems and Technology (TIST) 3(3):40

\bibitem[{Yang and Liu(2013)}]{yang2013quantitative}
Yang YH, Liu JY (2013) Quantitative study of music listening behavior in a
  social and affective context. IEEE Trans on Multimedia 15(6):1304--1315

\bibitem[{Yoshii et~al.(2008)Yoshii, Goto, Komatani, Ogata, and
  Okuno}]{yoshii2008efficientRecommender}
Yoshii K, Goto M, Komatani K, Ogata T, Okuno HG (2008) An efficient hybrid
  music recommender system using an incrementally trainable probabilistic
  generative model. IEEE Trans on Audio, Speech, and Language Processing
  16(2):435--447

\bibitem[{Zangerle et~al.(2012)Zangerle, Gassler, and
  Specht}]{zangerle2012exploiting}
Zangerle E, Gassler W, Specht G (2012) Exploiting twitter's collective
  knowledge for music recommendations. In: Proc. of the WWW'12 Workshop on
  'Making Sense of Microposts', Lyon, France, April 16, 2012, pp 14--17

\bibitem[{Zeghidour et~al.(2018)Zeghidour, Usunier, Synnaeve, Collobert, and
  Dupoux}]{zeghidour2018end}
Zeghidour N, Usunier N, Synnaeve G, Collobert R, Dupoux E (2018) End-to-end
  speech recognition from the raw waveform. In: Interspeech 2018, 19th Annual
  Conference of the International Speech Communication Association, Hyderabad,
  India, 2018, pp 781--785

\bibitem[{Zhou et~al.(2018)Zhou, Wang, Fang, Bui, and Berg}]{zhou2018visual}
Zhou Y, Wang Z, Fang C, Bui T, Berg TL (2018) Visual to sound: Generating
  natural sound for videos in the wild. In: Proc. of the IEEE Conference on
  Computer Vision and Pattern Recognition, pp 3550--3558

\end{thebibliography}
